\documentclass[journal]{IEEEtran}

\usepackage{amsmath,amssymb,bm,mathtools}
\usepackage{graphicx}

\usepackage{amsthm}
\theoremstyle{definition}
\newtheorem{myremark}{Remark}[section]
\newtheorem{myassumption}{Assumption}[section]
\newtheorem{myproposition}{Proposition}[section]
\newtheorem{myconjecture}{Conjecture}[section]

\newcommand{\minimize}{\operatornamewithlimits{\mathrm{minimize}}}
\newcommand{\argmin}{\operatornamewithlimits{\mathrm{arg\ min}}}
\newcommand{\plim}{\operatornamewithlimits{\mathrm{plim}}}
\newcommand{\calN}{\mathcal{N}}

\newcommand{\sign}{\mathrm{sign}}
\newcommand{\txv}{\text{v}}

\newcommand{\prox}{\mathrm{prox}}

\newcommand{\Pto}{\xrightarrow{\mathrm{P}}}
\newcommand{\abs}[1]{\left\lvert#1\right\rvert}
\newcommand{\norm}[1]{\left\lVert#1\right\rVert}
\newcommand{\Ex}[1]{\mathrm{E}\left[#1\right]}
\newcommand{\paren}[1]{\left(#1\right)}
\newcommand{\sqbra}[1]{\left[#1\right]}
\newcommand{\curbra}[1]{\left\{#1\right\}}

\allowdisplaybreaks[1]

\begin{document}
\title{Asymptotics of Proximity Operator for Squared Loss and Performance Prediction of Nonconvex Sparse Signal Recovery}
%
\author{
	Ryo~Hayakawa
	\thanks{
		This work was supported in part by JSPS KAKENHI Grant Number JP20K23324 and 24K17277. 
	}%
	\thanks{
		Ryo Hayakawa is with Institute of Engineering, Tokyo University of Agriculture and Technology, Tokyo, Japan (e-mail: hayakawa@go.tuat.ac.jp). 
	}
}%
\markboth{ACCEPTED TO APSIPA TRANSACTIONS ON SIGNAL AND INFORMATION PROCESSING}%
{R. Hayakawa: Asymptotics of Proximity Operator for Squared Loss and Performance Prediction of Nonconvex Sparse Signal Recovery}
%
\maketitle
\begin{abstract}
    Proximal splitting-based convex optimization is a promising approach to linear inverse problems because we can use some prior knowledge of the unknown variables explicitly. 
    An understanding of the behavior of the optimization algorithms would be important for the tuning of the parameters and the development of new algorithms. 
    In this paper, we first analyze the asymptotic property of the proximity operator for the squared loss function, which appears in the update equations of some proximal splitting methods for linear inverse problems. 
    Our analysis shows that the output of the proximity operator can be characterized with a scalar random variable in the large system limit. 
    Moreover, we apply the asymptotic result to the prediction of optimization algorithms for compressed sensing. 
    Simulation results demonstrate that the MSE performance of the Douglas-Rachford algorithm can be well predicted in compressed sensing with the $\ell_{1}$ optimization. 
    We also examine the behavior of the prediction for the case with nonconvex smoothly clipped absolute deviation (SCAD) and minimax concave penalty (MCP) regularization. 
\end{abstract}
%
\IEEEpeerreviewmaketitle
%
\section{Introduction} \label{sec:intro}
Linear inverse problems, i.e., the reconstruction of an unknown vector from its linear measurements, often appear in the field of signal processing. 
The linear inverse problem is called underdetermined when the number of measurements is less than that of elements of the unknown vector. 
In such underdetermined problems, we often use some information about the unknown vector to obtain a good reconstruction result. 
In compressed sensing~\cite{candes2005,candes2006,donoho2006,candes2008b}, for example, the unknown sparse vector can be effectively reconstructed by leveraging its sparsity. 
Another example of the underdetermined problems is the overloaded signal detection in wireless communications~\cite{sasahara2017,hayakawa2017a, hayakawa2018a}, where we can use the discreteness of the unknown vector. 

A common approach to underdetermined linear inverse problems is to solve an optimization problem using some regularizer based on the prior knowledge of the unknown vector. 
In compressed sensing, one of the most widely used optimization problems is the $\ell_{1}$ optimization. 
The objective function includes the $\ell_{1}$ regularization term to promote sparsity. 
Even though the objective function is not differentiable because of the $\ell_{1}$ norm, some proximal splitting methods can solve the problem with reasonable computational complexity. 
Examples of such methods include the iterative soft thresholding algorithm (ISTA)~\cite{daubechies2004,combettes2005,figueiredo2007} and the fast ISTA (FISTA)~\cite{beck2009}. 
Alternating direction method of multipliers (ADMM)~\cite{gabay1976,boyd2011} and the Douglas-Rachford algorithm~\cite{lions1979,eckstein1992,combettes2011} can both be used to solve the $\ell_{1}$ optimization problem. 
Despite the requirement of matrix inversion for linear inverse problems, ADMM and the Douglas-Rachford algorithm exhibit faster convergence than some gradient-based methods, such as ISTA, in certain cases~\cite{combettes2019,combettes2019a}. 

The asymptotic error performance of several optimization-based approaches for linear inverse problems has been analyzed by using some techniques~\cite{donoho2011,bayati2012,donoho2013}. 
Especially, the convex Gaussian min-max theorem (CGMT)~\cite{thrampoulidis2015,thrampoulidis2018} enables us to obtain the asymptotic error in a precise manner for the optimizer of various optimization problems. 
For instance, the asymptotic mean-square-error (MSE) of various regularized estimators has been analyzed in~\cite{atitallah2017,thrampoulidis2018}. 
In~\cite{thrampoulidis2018a,hayakawa2020a}, the symbol error rate has also been analyzed for the reconstruction of discrete-valued vectors. 

The above analyses have focused on the performance of the \emph{optimizer}, whereas the application of CGMT for the performance prediction of \emph{optimization algorithms} has been considered in~\cite{hayakawa2022}. 
In~\cite{hayakawa2022}, we have proposed a prediction method for the error performance of the tentative estimate at each iteration of ADMM. 
Such performance prediction could be utilized to understand the behavior of the algorithm, tune the parameter, and develop a new algorithm. 
For optimization algorithms other than ADMM, however, such asymptotic performance prediction has not been discussed for linear inverse problems in the literature. 
Expanding the prediction method to other optimization algorithms would be important for the development of new algorithms and the understanding of the behavior of the algorithms. 

In this paper, we investigate the asymptotic behavior of algorithms using the proximity operator of the squared loss function. 
As the target optimization problem, we consider the objective function with the squared loss function and a separable regularizer. 
Since the proximity operator of the squared loss function is not element-wise, we investigate its asymptotic property via the CGMT framework. 
Our main finding is that the optimization problem in the definition of the proximity operator can be analyzed by using CGMT for the squared loss function. 
Specifically, we show that the output of the proximity operator for the squared loss function can be regarded as a scalar random variable in the large system limit. 

By using the analytical result, we also propose a prediction method for the evolution of the asymptotic MSE in proximal splitting methods  composed of the proximity operator of the squared loss function and a separable regularizer. 
As an representative example of such algorithms, we here consider the Douglas-Rachford algorithm in this paper.
While the derivation of the proposed method is partly non-rigorous, their simulation results for the $\ell_{1}$ optimization demonstrate their potential as a foundation for more rigorous discussion in future work. 
These findings would contribute to the broader understanding of optimization algorithms in high-dimensional settings. 
Moreover, we also investigate the behavior for the case with nonconvex regularization in the simulation, though only the convex optimization is considered in~\cite{hayakawa2022}. 
The effectiveness of the performance prediction across various scenarios highlights its potential to extend to other optimization algorithms, including those involving nonconvex cases.

We use the following notations in this paper. 
The transpose and the identity matrix are denoted by $(\cdot)^{\top}$ and $\bm{I}$, respectively. 
The $\ell_{1}$ norm and the $\ell_{2}$ norm for a vector $\bm{u}=\sqbra{u_{1}\, \cdots \, u_{N}}^{\top} \in \mathbb{R}^{N}$ are defined as $\norm{\bm{u}}_{1} = \sum_{n=1}^{N} \abs{u_{n}}$ and $\norm{\bm{u}}_{2} = \sqrt{\sum_{n=1}^{N} u_{n}^{2}}$, respectively. 
We denote the sign function by $\sign(\cdot)$. 
For a lower semicontinuous function $\phi: \mathbb{R}^{N} \to \mathbb{R} \cup \curbra{+\infty}$, the proximity operator is defined as $\prox_{\phi} (\bm{u}) = \argmin_{\bm{s} \in \mathbb{R}^{N}} \curbra{ \phi(\bm{s}) + \frac{1}{2} \norm{\bm{s}-\bm{u}}_{2}^{2}}$. 
We denote the Gaussian distribution with mean $\mu$ and variance $\sigma^{2}$ by $\mathcal{N}(\mu, \sigma^{2})$. 
If random variables $\curbra{\Theta_{n}}$ ($n=1,2,\dotsc$) with index $n$ converge in probability to $\Theta$, we use the notation $\Theta_{n} \Pto \Theta$ as $n \to \infty$ or $\plim_{n \to \infty} \Theta_{n} = \Theta$. 
%
\section{Optimization for Linear Inverse Problems} \label{sec:problem}
\subsection{Linear Inverse Problems}
In this paper, we discuss the linear inverse problem, which means the reconstruction of a vector $\bm{x} \in \mathbb{R}^{N}$ from its measurement $\bm{y} \in \mathbb{R}^{M}$ given by 
\begin{align}
	\bm{y} = \bm{A} \bm{x} + \bm{v}. \label{eq:measurement}
\end{align}
The measurement matrix $\bm{A} \in \mathbb{R}^{M \times N}$ is assumed to be known and $\bm{v} \in \mathbb{R}^{M}$ denotes a noise vector. 
We mainly focus on the underdetermined case where the measurement ratio $\Delta \coloneqq M / N$ is less than 1 and the solution is not unique even in the noiseless case. 
A common approach in this situation is the utilization of the structure of $\bm{x}$. 
In compressed sensing, for example, the unknown vector is assumed to be sparse, i.e., the vector has many zero elements. 
In some applications in wireless communications, the unknown vector has other structures such as boundedness and discreteness~\cite{tan2001, nagahara2015, aissa-el-bey2015, hayakawa2018b}. 

\subsection{Optimization-Based Approach}
Mathematical optimization is a useful method for underdetermined linear inverse problems. 
In this approach, we can design the objective function of the problem to make the most of the structure of the unknown vector $\bm{x}$. 
As a simple example, we discuss the following optimization problem 
\begin{align}
    \minimize_{\bm{s} \in \mathbb{R}^{N}}\ 
    \curbra{
        L(\bm{s}) 
        + f_{\lambda}({\bm{s}}) 
    } \label{eq:optimization}
\end{align}
in this paper. 
We refer to $L(\bm{s}) = \frac{1}{2} \norm{ \bm{y} - \bm{A} \bm{s} }_{2}^{2}$ as the squared loss function. 
The function $f_{\lambda} (\cdot): \mathbb{R}^{N} \to \mathbb{R} \cup \curbra{+\infty}$ is a regularizer that can incorporate prior knowledge of the unknown vector $\bm{x}$. 
The parameter $\lambda$ determines the balance between two terms in the objective function. 

In the sparse vector reconstruction, for example, the $\ell_{1}$ regularization by $f_{\lambda} (\bm{s}) = \lambda \norm{\bm{s}}_{1}$ ($\lambda > 0$) is widely used as a convex regularizer. 
The elastic net~\cite{Zou2005-pr} given by $f_{\lambda} (\bm{s}) = \lambda_{1} \norm{\bm{s}}_{1} + \frac{\lambda_{2}}{2} \norm{\bm{s}}_{2}^{2}$, where $\lambda_{1}, \lambda_{2}$ ($>0$) are the parameters, is also a popular convex regularizer. 
Other than these convex regularizers, various nonconvex regularizers have been proposed for the sparse vector reconstruction~\cite{wen2018}. 
For example, the smoothly clipped absolute deviation (SCAD)~\cite{fan2001} regularizer is given by $f_{\lambda} (\bm{s}) = \sum_{n=1}^{N} \tilde{f}_{\lambda} (s_{n})$, where  
\begin{align}
    \tilde{f}_{\lambda} (s) 
    &= 
    \begin{cases}
        \lambda \abs{s} & (\abs{s} \le \lambda) \\[5pt]
        \dfrac{2 a \lambda \abs{s} - s^{2} - \lambda^{2}}{2 (a - 1)} & (\lambda < \abs{s} \le a \lambda) \\[10pt]
        \dfrac{(a + 1) \lambda^{2}}{2} & (\abs{s} > a \lambda) 
    \end{cases} \label{eq:SCAD}
\end{align}
and $a$ ($ > 1$) is the parameter. 
For the minimax concave penalty (MCP)~\cite{zhang2010a} regularizer, the function $\tilde{f}_{\lambda} (s)$ is given by
\begin{align}
    \tilde{f}_{\lambda} (s) 
    &= 
    \begin{cases}
        \lambda \abs{s} - \dfrac{s^{2}}{2 b} & (\abs{s} \le b \lambda) \\[5pt]
        \dfrac{b \lambda^{2}}{2} & (\abs{s} > b \lambda) 
    \end{cases}, \label{eq:MCP}
\end{align}
where $b$ ($ > 0$) is the parameter. 

Various optimization algorithms have been proposed for the optimization problem in~\eqref{eq:optimization}. 
For example, the Douglas-Rachford algorithm~\cite{lions1979,eckstein1992,combettes2011} solves the optimization problem in~\eqref{eq:optimization} by using the proximity operators of $L(\cdot)$ and $f_{\lambda} (\cdot)$. 
The update equations of the algorithm with the iteration index $k$ ($= 0, 1, 2, \dotsc$) can be written as 
\begin{align}
    \bm{s}^{(k+1)} 
    &= 
    \prox_{\gamma L} \paren{\bm{z}^{(k)}}, \label{eq:update_s_DR} \\
    \bm{z}^{(k+1)} 
    &= 
    \bm{z}^{(k)} + \rho_{k} \paren{\prox_{\gamma f_{\lambda}} \paren{2 \bm{s}^{(k + 1)} - \bm{z}^{(k)}} - \bm{s}^{(k+1)}}, \label{eq:update_z_DR}
\end{align}
where $\gamma$ ($>0$) and $\rho_{k} \in [\varepsilon, 2 - \varepsilon]$ ($\varepsilon \in (0, 1)$) are the parameters in the algorithm. 
By definition, we can obtain the proximity operator of the function $L(\cdot)$ as 
\begin{align}
    \hspace{-1mm}
    \prox_{\gamma L} \paren{\bm{z}} 
    &= 
    \argmin_{\bm{s} \in \mathbb{R}^{N}} 
    \curbra{\frac{1}{2} \norm{ \bm{y} - \bm{A} \bm{s} }_{2}^{2} + \frac{1}{2 \gamma} \norm{\bm{s} - \bm{z}}_{2}^{2}} \label{eq:prox_L_def} \\
    &= 
    \paren{\bm{A}^{\top} \bm{A} + \frac{1}{\gamma} \bm{I}}^{-1} \paren{\bm{A}^{\top} \bm{y} + \frac{1}{\gamma} \bm{z}}, \label{eq:prox_L}
\end{align}
where $\bm{z} \in \mathbb{R}^{N}$ and $\gamma > 0$. 
The proximity operator of $f_{\lambda} (\cdot)$ can also be computed efficiently for various regularizers. 
When $f_{\lambda} (\bm{s}) = \lambda \norm{\bm{s}}_{1}$, for example, the proximity operator of the function $f_{\lambda}(\cdot)$ can be written as 
\begin{align}
    \sqbra{\prox_{\gamma f_{\lambda}} (\bm{r})}_{n}
    &= 
    \sign (r_{n}) \max(\abs{r_{n}} - \gamma \lambda, 0), \label{eq:prox_L1}
\end{align} 
where $r_{n}$ and $\sqbra{\prox_{\gamma f_{\lambda}} (\bm{r})}_{n}$ are the $n$-th element of $\bm{r}$ and $\prox_{\gamma f_{\lambda}} (\bm{r})$, respectively. 
For the elastic net regularizer $f_{\lambda} (\bm{s}) = \lambda_{1} \norm{\bm{s}}_{1} + \frac{\lambda_{2}}{2} \norm{\bm{s}}_{2}^{2}$, the proximity operator can be computed as
\begin{align}
    \sqbra{\prox_{\gamma f_{\lambda}} (\bm{r})}_{n}
    &= 
    \dfrac{\sign (r_{n}) \max(\abs{r_{n}} - \gamma \lambda_{1}, 0)}{1 + \gamma \lambda_{2}}. \label{eq:prox_EN}
\end{align}
As for the SCAD regularizer in~\eqref{eq:SCAD}, the proximity operator is given by 
\begin{align}
    &\sqbra{\prox_{\gamma f_{\lambda}} (\bm{r})}_{n} \notag \\
    &= 
    \begin{cases}
        \sign (r_{n}) \max(\abs{r_{n}} - \gamma \lambda, 0) & (\abs{r_{n}} \le (1 + \gamma) \lambda) \\[5pt]
        \dfrac{(a - 1) r_{n} -  \sign (r_{n}) a \gamma \lambda}{a - 1 - \gamma} & ((1 + \gamma) \lambda < \abs{r_{n}} \le a \lambda) \\[5pt]
        r_{n} & (\abs{r_{n}} > a \lambda)
    \end{cases}. \label{eq:prox_SCAD1}
\end{align}
for $a > 1 + \gamma$. 
Similarly, the proximity operator of the MCP regularizer in~\eqref{eq:MCP} can be computed as 
\begin{align}
    &\sqbra{\prox_{\gamma f_{\lambda}} (\bm{r})}_{n} \notag \\
    &= 
    \begin{cases}
        0 & (\abs{r_{n}} \le \gamma \lambda) \\[5pt]
        \dfrac{b}{b - \gamma} (r_{n} - \sign(r_{n}) \gamma \lambda) & (\gamma \lambda < \abs{r_{n}} \le b \lambda) \\[5pt]
        r_{n} & (\abs{r_{n}} > b \lambda)
    \end{cases} \label{eq:prox_MCP1}
\end{align}
for $b > \gamma$. 
By computing~\eqref{eq:update_s_DR} and~\eqref{eq:update_z_DR} iteratively, we can obtain a sequence $\curbra{\bm{s}^{(k)}}_{k = 1, 2, \dotsc}$ converging to the solution of the optimization problem in~\eqref{eq:optimization} when the regularizer is convex. 
When the regularizer is nonconvex, the convergence of the sequence is not necessarily guaranteed.
%
\section{Main Results} \label{sec:result}
\subsection{Asymptotics of Proximity Operator for Squared Loss}
We firstly analyze the output of the proximity operator $\prox_{\gamma L} (\cdot)$ in~\eqref{eq:prox_L}. 
In our analysis, we assume the large system limit $M, N \to \infty$ ($M / N = \Delta$), where the sequence of problems with $\curbra{\bm{x}, \bm{A}, \bm{v}}$ indexed by $N$ is considered as in several high-dimensional analyses~\cite{thrampoulidis2018}. 
We also assume the following conditions as in~\cite{hayakawa2022}. 
\begin{myassumption} \label{ass:distribution}
    The elements of the vector $\bm{x} \in \mathbb{R}^{N}$ are independent and identically distributed (i.i.d.). 
    The distribution $p_{X}$ of $\bm{x}$ is known and has some finite mean and variance. 
    The measurement matrix $\bm{A} \in \mathbb{R}^{M \times N}$ has i.i.d.\ random variables with $\mathcal{N}(0, 1/N)$. 
    The noise vector $\bm{v} \in \mathbb{R}^{M}$ has i.i.d.\ random variables with $\mathcal{N}(0, \sigma_{\txv}^{2})$. 
\end{myassumption}

\begin{myremark}
    We assume that each element of the measurement matrix $\bm{A}$ follows a Gaussian distribution in Assumption~\ref{ass:distribution}. 
    This is because we require the Gaussian assumption in the analysis with CGMT~\cite{thrampoulidis2018}. 
    Nonetheless, the universality of random matrices discussed in~\cite{bayati2015,panahi2017,oymak2018} implies that the analytical result still holds for some other distributions. 
    In~\cite[Section VIII-F]{thrampoulidis2018}, for example, it is expected that the analysis via CGMT holds for i.i.d.\ sub-Gaussian matrix. 
    Computer simulations in~\cite{thrampoulidis2018,hayakawa2022} show that the empirical performance is well predicted even when the measurement matrix is composed of i.i.d.\ Bernoulli distribution with $p = 0.5$. 
\end{myremark}

Under Assumption~\ref{ass:distribution}, we can obtain the following result on the asymptotic behavior of the proximity operator $\prox_{\gamma L} (\cdot)$. 
\begin{myproposition} \label{prop:prox}
    It is assumed that $\bm{x}$, $\bm{A}$, and $\bm{v}$ satisfy the conditions in Assumption~\ref{ass:distribution}. 
    We then consider the output of the proximity operator given by
    \begin{align}
        \hat{\bm{s}} 
        &= 
        \prox_{\gamma L} \paren{\bm{z}}, \label{eq:s_hat}
    \end{align}
    where $\bm{z}$ has i.i.d.\ elements with a distribution $p_{Z}(z)$ and is assumed to be independent of $\bm{A}$. 
    Here, we assume that the optimization problem 
    \begin{align}
        &\min_{\alpha > 0} \max_{\beta \ge 0} 
        \curbra{
            \frac{\alpha\beta\sqrt{\Delta}}{2} 
            + \frac{\beta \sigma_{\txv}^{2} \sqrt{\Delta}}{2\alpha} 
            - \frac{1}{2} \beta^{2} 
            + \Ex{J (\alpha, \beta; Z)} 
        } \label{eq:SO}
    \end{align}
    has a unique optimizer $(\alpha^{\ast}, \beta^{\ast})$\footnote{The uniqueness of the solution can be established under certain conditions (e.g., the boundedness of the set of minimizers). However, in the general case, completely removing this assumption proves to be challenging. For further details, please refer to~\cite[Remark 19]{thrampoulidis2018}.}. 
    In~\eqref{eq:SO}, 
    \begin{align}
        J (\alpha, \beta; Z) 
        &= 
        \frac{\beta\sqrt{\Delta}}{2\alpha} \paren{\hat{S} (\alpha, \beta; Z) - X}^{2} \notag \\
        &\hspace{4mm} 
        - \beta H \paren{\hat{S} (\alpha, \beta; Z) - X} \notag \\
        &\hspace{4mm} 
        +\frac{1}{2 \gamma} \paren{ \hat{S} (\alpha, \beta; Z) - Z}^{2}, \label{eq:J_alpha_beta} \\
        \hat{S} (\alpha, \beta; Z) 
        &= 
        \dfrac{1}{\dfrac{\beta\sqrt{\Delta}}{\alpha} + \dfrac{1}{\gamma}} 
        \paren{ 
            \frac{\beta\sqrt{\Delta}}{\alpha} 
            \paren{X + \dfrac{\alpha}{\sqrt{\Delta}} H}
            + \frac{1}{\gamma} Z
        }, \label{eq:S_alpha_beta} 
    \end{align}
    $X \sim p_{X}$, $H \sim \mathcal{N}(0, 1)$, and $Z \sim p_{Z}$. 
    The expectation $\Ex{\cdot}$ in~\eqref{eq:SO} is calculated over all random variables $X$, $H$, and $Z$. 
    Then, the following statements hold: 
    \begin{enumerate}
        \item 
            The asymptotic MSE of $\hat{\bm{s}}$ in~\eqref{eq:s_hat} can be written as 
            \begin{align}
                \plim_{M, N \to \infty} 
                \frac{1}{N} \norm{\hat{\bm{s}} - \bm{x}}_{2}^{2} 
                &= 
                \paren{\alpha^{\ast}}^{2} - \sigma_{\txv}^{2}. \label{eq:MSE_prox}
            \end{align}
        \item 
            Let $\mu_{\hat{\bm{s}}}$ denote the empirical distribution of $\hat{\bm{s}} = \sqbra{\hat{s}_{1}\ \dotsb\ \hat{s}_{N}}^{\top} \in \mathbb{R}^{N}$ that corresponds to the cumulative distribution function (CDF) given by $P_{\hat{\bm{s}}} (s) = \frac{1}{N} \sum_{n=1}^{N} \mathbb{I}\paren{\hat{s}_{n} < s}$, where $\mathbb{I}\paren{\hat{s}_{n} < s} = 1$ if $\hat{s}_{n} < s$ and otherwise $\mathbb{I}\paren{\hat{s}_{n} < s} = 0$. 
            Then, the distribution $\mu_{\hat{\bm{s}}}$ converges weakly in probability to the distribution $\mu_{S}$ of $S = \hat{S} \paren{\alpha^{\ast}, \beta^{\ast}; Z}$, i.e., $\int g d\mu_{\hat{\bm{s}}} \Pto \int g d\mu_{S}$ holds for any continuous compactly supported function $g(\cdot): \mathbb{R} \to \mathbb{R}$. 
    \end{enumerate}
\end{myproposition}
\begin{proof}[Sketch of proof]
    By definition in~\eqref{eq:s_hat}, $\hat{\bm{s}}$ is characterized as the solution of the optimization problem in~\eqref{eq:prox_L_def}. 
    Similarly to the discussion in~\cite[Remark~IV.1]{hayakawa2020a}, we can prove~\eqref{eq:MSE_prox} by using the standard method with CGMT~\cite{thrampoulidis2018}. 
    Moreover, the second statement of Proposition~\ref{prop:prox} is proven in the procedure of~\cite[Theorem~III.2]{hayakawa2020a}. 
    Hence, we do not include the details of the rigorous proof. 
    For the overview of the derivation of~\eqref{eq:SO}, see Appendix~\ref{app:proof}. 
\end{proof}

The first statement in Proposition~\ref{prop:prox} means that the asymptotic MSE for the output of the proximity operator can be predicted by solving the scalar optimization problem in~\eqref{eq:SO}. 
The second one implies that the distribution of the elements of the vector $\hat{\bm{s}}$ can be characterized by the random variable $S = \hat{S} \paren{\alpha^{\ast}, \beta^{\ast}; Z}$. 
We can thus consider $\hat{S}\paren{\alpha^{\ast}, \beta^{\ast}; Z}$ in~\eqref{eq:S_alpha_beta} as a decoupled version of the proximity operator in~\eqref{eq:s_hat} intuitively. 
For a similar discussion, see~\cite[Fig.~1]{hayakawa2022}. 

The optimization of $\alpha$ and $\beta$ in~\eqref{eq:SO} can be performed by using searching methods like the golden section search~\cite{luenberger2008}. 
Since it is difficult to compute the expectation in~\eqref{eq:SO} exactly in general, we need to approximate it with the average of many realizations of $X$, $H$, and $Z$. 
\subsection{Application to Performance Prediction of Optimization Algorithm}
Although the proximity operator itself is not a valid reconstruction method in general, we believe that the result in Proposition~\ref{prop:prox} is not only of theoretical interest but also serves as a foundation for understanding and designing optimization algorithms in linear inverse problems. 
By using Proposition~\ref{prop:prox}, for example, 
we can derive an error prediction method for the estimate $\bm{s}^{(k)}$ in some optimization algorithm. 
To obtain the prediction method, we make an additional assumption that the regularizer $f_{\lambda} (\cdot)$ is convex and separable as follows. 
\begin{myassumption} \label{ass:regularizer}
    The function $f_{\lambda} (\cdot): \mathbb{R}^{N} \to \mathbb{R} \cup \curbra{+\infty}$ is separable and can be written as $f_{\lambda} (\bm{s}) = \sum_{n=1}^{N} \tilde{f}_{\lambda} (s_{n})$ with a convex function $\tilde{f}_{\lambda} (\cdot): \mathbb{R} \to \mathbb{R} \cup \curbra{+\infty}$. 
    We sometimes use the notation $f_{\lambda} (\cdot)$ for the corresponding function $\tilde{f}_{\lambda} (\cdot)$ with the slight abuse of notation. 
\end{myassumption}
\begin{myremark}
    For the separable regularizer $f_{\lambda} (\cdot)$ in Assumption~\ref{ass:regularizer}, the proximity operator $\prox_{\gamma f_{\lambda}} (\cdot): \mathbb{R}^{N} \to \mathbb{R}^{N}$ becomes an element-wise function. 
    In other words, the $n$-th element of the output of the proximity operator is the function of the $n$-th element of the input only. 
\end{myremark}

To predict the performance of an algorithm by using Proposition~\ref{prop:prox}, it is necessary that the update equations except for $\prox_{\gamma L}(\cdot)$ are element-wise. 
As a representative example of such algorithms, 
we derive the conjecture below for Douglas-Rachford algorithm in~\eqref{eq:update_s_DR} and~\eqref{eq:update_z_DR}. 
Note that the derivation of the conjecture is partly non-rigorous as~\cite[Claim~III.1]{hayakawa2022}. 
\begin{myconjecture} \label{conj:DR}
    Suppose that Assumptions~\ref{ass:distribution} and~\ref{ass:regularizer} hold. 
    We define the stochastic process 
    \begin{align}
        S_{k+1} 
        &= 
        \hat{S}\paren{\alpha_{k}^{\ast}, \beta_{k}^{\ast}; Z_{k}} \label{eq:update_S} \\
        Z_{k+1} 
        &= 
        Z_{k} + \rho_{k} \paren{\prox_{\gamma f_{\lambda}} \paren{2S_{k+1} - Z_{k}} - S_{k+1}} \label{eq:update_Z} 
    \end{align}
    with the index $k$. 
    We here assume that the optimization problem 
    \begin{align}
        &\min_{\alpha > 0} \max_{\beta \ge 0} 
        \curbra{
            \frac{\alpha\beta\sqrt{\Delta}}{2} 
            + \frac{\beta \sigma_{\txv}^{2} \sqrt{\Delta}}{2\alpha} 
            - \frac{1}{2} \beta^{2} 
            + \Ex{J(\alpha, \beta; Z_{k})} 
        } \label{eq:SO_k}
    \end{align}
    has a unique optimizer $(\alpha_{k}^{\ast}, \beta_{k}^{\ast})$. 
    Then, we have the following conjecture: 
    \begin{enumerate}
        \item 
            The asymptotic MSE of $\bm{s}^{(k+1)}$ in~\eqref{eq:update_s_DR} can be written as 
            \begin{align}
                \plim_{M, N \to \infty} 
                \frac{1}{N} \norm{\bm{s}^{(k+1)} - \bm{x}}_{2}^{2} 
                &= 
                \paren{\alpha_{k}^{\ast}}^{2} - \sigma_{\txv}^{2}. \label{eq:MSE_DR}
            \end{align}
        \item The empirical distribution of $\bm{s}^{(k+1)}$ converges weakly in probability to the distribution of $S_{k+1}$. 
    \end{enumerate}
\end{myconjecture}
\begin{proof}[Outline of derivation]
Since $\bm{z}^{(k)}$ in~\eqref{eq:update_s_DR} and~\eqref{eq:update_z_DR} is not independent of $\bm{A}$, Proposition~\ref{prop:prox} cannot be used to~\eqref{eq:update_s_DR} directly. 
We thus take the same procedure as the approach in~\cite[Claim~III.1]{hayakawa2022}, where an unproven extension of CGMT has been assumed for the performance prediction of iterative algorithms. 
Under the assumption, when the element of $\bm{z}^{(k)}$ can be regarded as the random variable $Z_{k}$, the distribution of the element of $\bm{s}^{(k+1)}$ in~\eqref{eq:update_s_DR} can be described by the distribution of the random variable $S_{k+1}$ in~\eqref{eq:update_S} as in Proposition~\ref{prop:prox}. 
As for~\eqref{eq:update_z_DR}, the update of $\bm{z}^{(k)}$ can be decoupled into~\eqref{eq:update_Z} because the update is element-wise under Assumption~\ref{ass:regularizer}. 
Since the procedure of the derivation is the same as~\cite[Claim~III.1]{hayakawa2022}, the detailed explanation for the derivation is omitted. 
\end{proof}

Conjecture~\ref{conj:DR} implies that the evolution of the asymptotic MSE in the Douglas-Rachford algorithm can be predicted by solving the optimization problem in~\eqref{eq:SO_k} and computing~\eqref{eq:MSE_DR} for each $k$. 
The updates of $S_{k}$ and $Z_{k}$ in~\eqref{eq:update_S} and~\eqref{eq:update_Z} can be seen as a simplified version of the update of $\bm{s}^{(k)}$ and $\bm{z}^{(k)}$ in~\eqref{eq:update_s_DR} and~\eqref{eq:update_z_DR}, respectively. 

From the result in Conjecture~\ref{conj:DR}, we can tune the parameters in the Douglas-Rachford algorithm for fast convergence. 
Since the relation between the parameters and the MSE is complicated, it is difficult to obtain the explicit expression of the optimal parameters. 
However, the result enables us to predict the error performance of the algorithm numerically and to select the parameter that achieves fast convergence in the asymptotic regime, without the empirical reconstruction. 

\begin{myremark}
    As discussed in~\cite{eckstein1992}, the Douglas-Rachford algorithm is closely related to ADMM analyzed in~\cite{hayakawa2022}. 
    In fact, the update equations of ADMM can be regarded as the Douglas-Rachford algorithm for the dual problem of the original optimization problem. 
    However, the relaxation parameters $\rho_{k}$ are included in the Douglas-Rachford algorithm considered in this paper, whereas the corresponding parameters are not considered in the analysis of~\cite{hayakawa2022}. 
    These parameters provide additional flexibility in tuning the algorithm for faster convergence. 
    Furthermore, since we have expressed Proposition~\ref{prop:prox} as the result for the proximity operator for the squared loss function, our results have the potential to be extended to other optimization algorithms that utilize the proximity operator.
\end{myremark}
%
\section{Simulation Results} \label{sec:simulation}
In this section, we demonstrate the validity of our approach via computer simulations. 
The distribution of an unknown vector $\bm{x}$ is assumed to be the Bernoulli-Gaussian distribution given by 
\begin{align}
    p_{X}(x) 
    &= 
    p_{0} \delta_{0}(x) + (1 - p_{0}) p_{H}(x). \label{eq:distribution_x}
\end{align}
Here, $\delta_{0}(\cdot)$ is the Dirac delta function, $p_{H}(\cdot)$ is the probability density function corresponding to $\mathcal{N}(0, 1)$, and $p_{0} \in (0, 1)$ represents the probability of $0$. 
The unknown vector $\bm{x}$ is sparse when $p_{0}$ is large. 
For the simulations, we assume that the measurement matrix $\bm{A}$ and the noise vector $\bm{v}$ satisfy the conditions in Assumption~\ref{ass:distribution} unless otherwise stated. 
For the optimization of $\alpha$ and $\beta$ in~\eqref{eq:SO} and~\eqref{eq:SO_k}, the golden section search is used. 
\subsection{MSE of Output of Proximity Operator for Squared Loss}
\begin{figure}[!t]
    \centering
    \includegraphics[width=85mm]{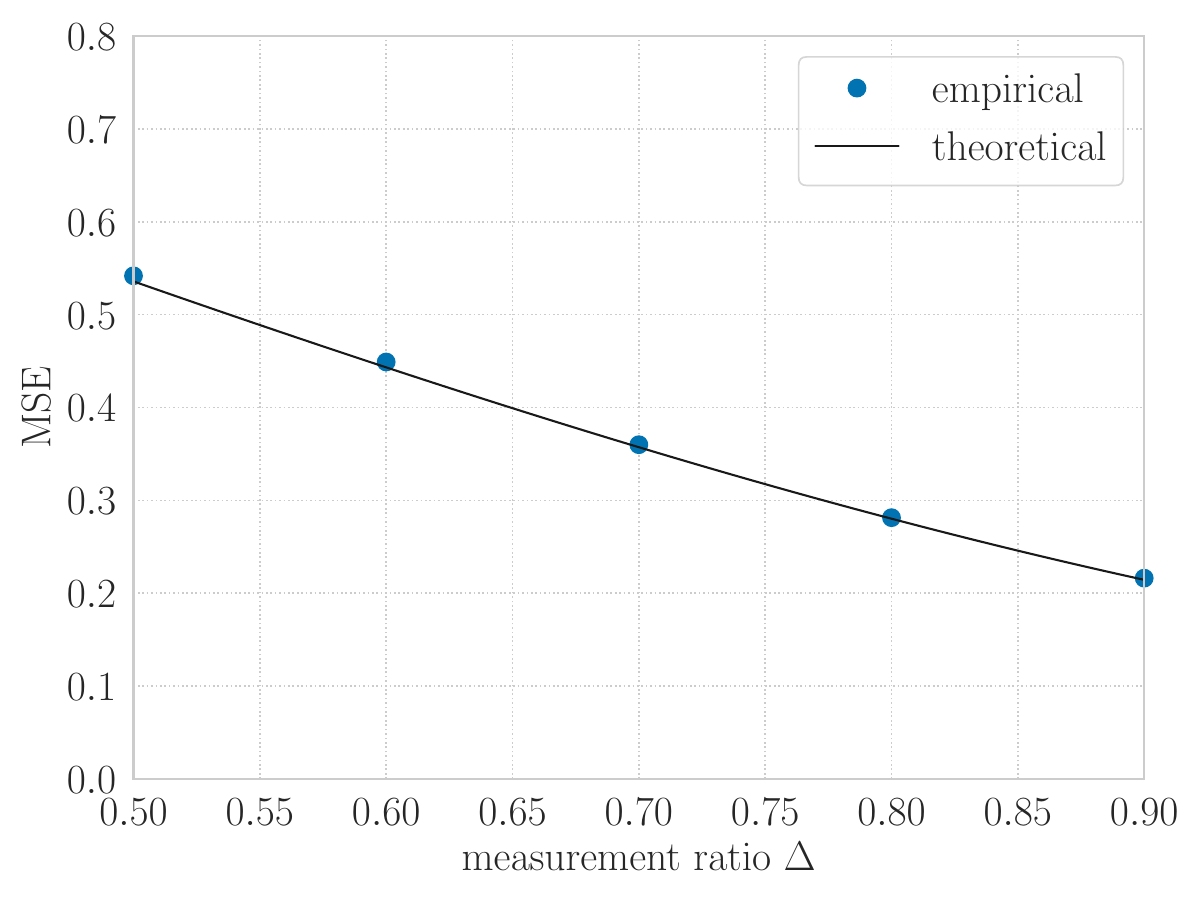}
    \caption{MSE of output of proximity operator for squared loss ($N = 1000$, $p_{0} = 0.95$, $\sigma_{\mathrm{v}}^{2} = 0.001$, $\gamma = 10$).}
    \label{fig:MSE_prox}
\end{figure}

We demonstrate a simulation result for the first statement of Proposition~\ref{prop:prox}. 
In Fig.~\ref{fig:MSE_prox}, the empirical MSE of $\hat{\bm{s}} = \prox_{\gamma L}(\bm{z})$ in~\eqref{eq:s_hat} is compared with the corresponding theoretical prediction $\paren{\alpha^{\ast}}^{2} - \sigma_{\txv}^{2}$ in~\eqref{eq:MSE_prox}. 
In the simulation, we set $N = 1000$, $p_{0} = 0.95$, $\sigma_{\mathrm{v}}^{2} = 0.001$, and $\gamma = 10$. 
In the figure, `empirical' denotes the empirical MSE of $\hat{\bm{s}}$ and `theoretical' denotes the corresponding asymptotic value $\paren{\alpha^{\ast}}^{2} - \sigma_{\txv}^{2}$. 
For the computation of the empirical MSE, we first make the realizations of $\bm{x}$, $\bm{A}$, and $\bm{v}$ to compute $\bm{y}$ from~\eqref{eq:measurement}. 
We then create each element of $\bm{z}$ from the i.i.d.\ standard Gaussian distribution. 
Using these realizations, we compute the MSE of $\hat{\bm{s}}$ from~\eqref{eq:prox_L} and~\eqref{eq:s_hat}. 
The empirical performance in the figure is obtained by averaging the MSE for $100$ independent realizations. 
For the theoretical prediction, we solve the scalar optimization problem in~\eqref{eq:SO} to obtain $\alpha^{\ast}$. 
Note that we approximate the expectation in the optimization problem with $100,000$ realizations of $X \sim p_{X}$, $H \sim \calN (0, 1)$ and $Z \sim \calN (0, 1)$. 
From Fig.~\ref{fig:MSE_prox}, we observe that the empirical performance and its theoretical prediction agree well with each other for various measurement ratios $\Delta$. 
\subsection{Distribution of Output of Proximity Operator for Squared Loss}
\begin{figure}[!t]
    \centering
    \includegraphics[width=85mm]{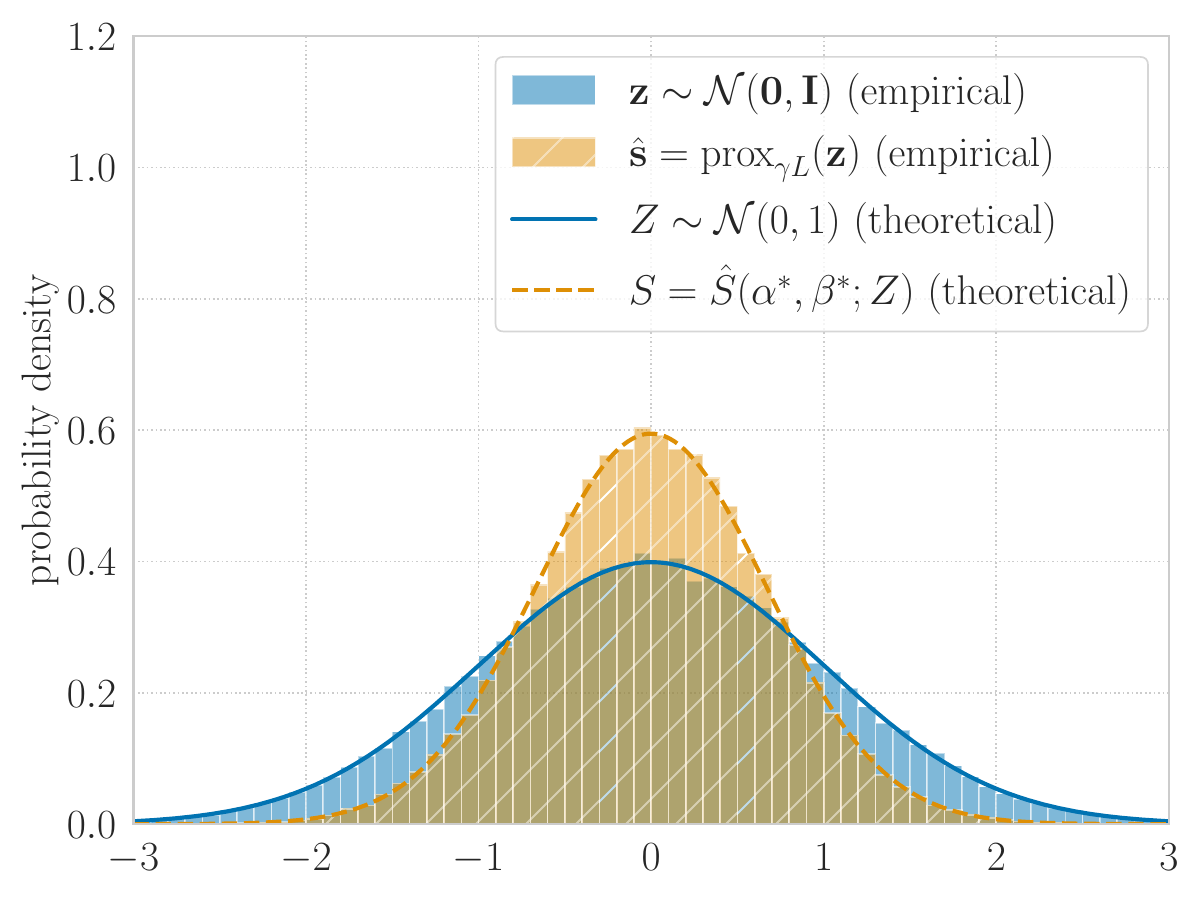}
    \caption{Comparison of probability density for $\hat{\bm{s}}$ and $S$ ($N = 1000$, $\Delta = 0.6$, $p_{0} = 0.95$, $\sigma_{\mathrm{v}}^{2} = 0.01$, $\gamma = 10$).}
    \label{fig:density}
\end{figure}

We then examine the second statement of Proposition~\ref{prop:prox}. 
In Fig~\ref{fig:density}, we compare the probability density of the elements of $\hat{\bm{s}} = \prox_{\gamma L}(\bm{z})$ with the theoretical result. 
In the simulation, we set $N = 1000$, $\Delta = 0.6$, $p_{0} = 0.95$, $\sigma_{\mathrm{v}}^{2} = 0.01$, and $\gamma = 10$. 
The empirical $\hat{\bm{s}} = \prox_{\gamma L}(\bm{z})$ and the realizations of the corresponding random variable $S = \hat{S} \paren{\alpha^{\ast}, \beta^{\ast}; Z}$ is obtained in the same way as the first experiment. 
The empirical histogram is obtained from the result of $100$ independent trials. 
From Fig.~\ref{fig:density}, we can see that the empirical distribution of $\hat{\bm{s}}$ agrees well with the distribution of $S$ obtained by Proposition~\ref{prop:prox}. 
\subsection{Performance Prediction of Douglas-Rachford Algorithm with Convex Regularization}
Next, we investigate the validity of Conjecture~\ref{conj:DR}. 
For the reconstruction of the sparse vector following~\eqref{eq:distribution_x}, we first consider the $\ell_{1}$ optimization with $f_{\lambda} (\bm{s}) = \lambda \norm{\bm{s}}_{1}$, i.e., 
\begin{align}
    \minimize_{\bm{s} \in \mathbb{R}^{N}} 
    \curbra{
        \frac{1}{2} \norm{ \bm{y}-\bm{A}\bm{s} }_{2}^{2} 
        + \lambda \norm{\bm{s}}_{1} 
    }, \label{eq:L1_optimization}
\end{align}
which is a widely used convex optimization problem in compressed sensing and satisfies Assumption~\ref{ass:regularizer}. 
For the $\ell_{1}$ regularization, the proximity operator is given by~\eqref{eq:prox_L1}. 

\begin{figure}[!t]
    \centering
    \includegraphics[width=85mm]{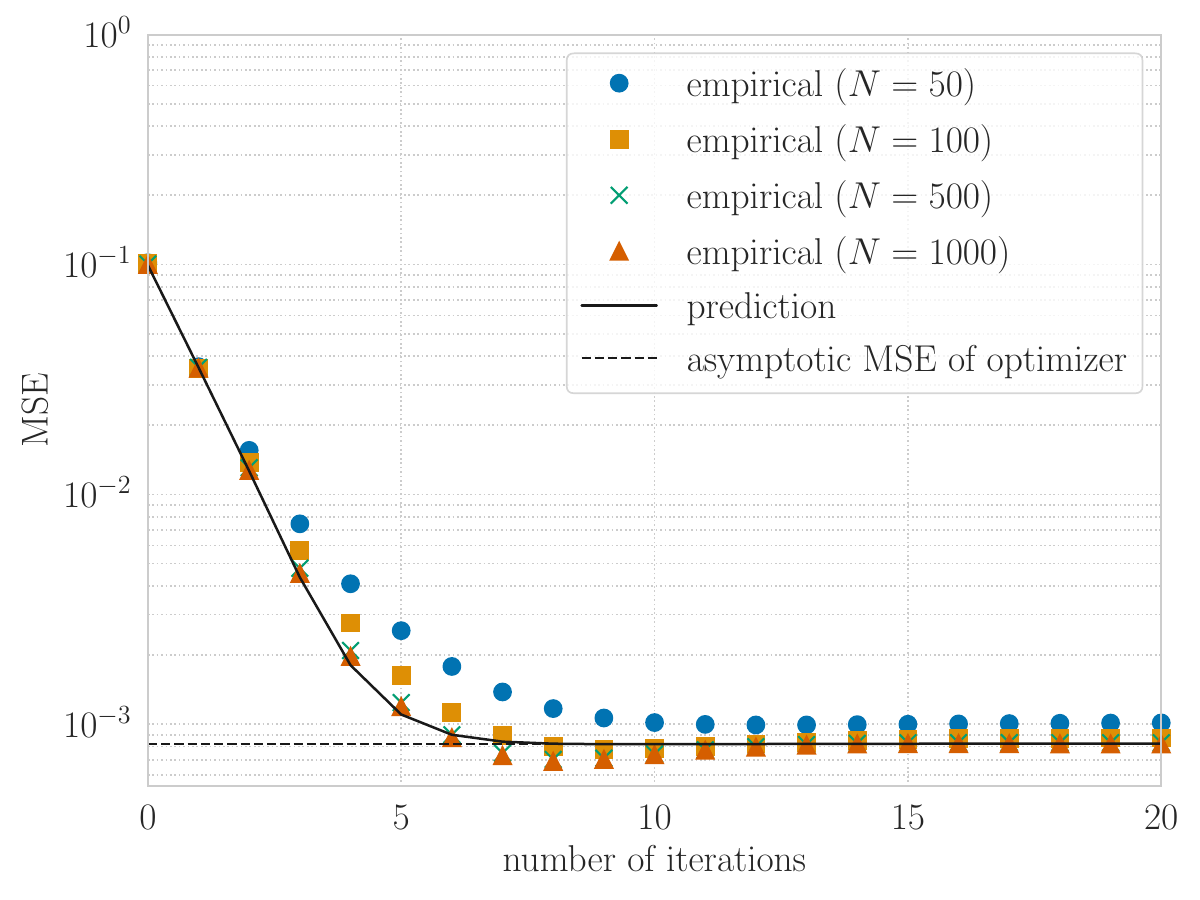}
    \caption{MSE performance of Douglas-Rachford algorithm for $\ell_{1}$ optimization ($\Delta = 0.7$, $p_{0} = 0.9$, $\sigma_{\txv}^{2} = 0.001$, $\gamma = 10$, $\rho_{k} = 1$).}
    \label{fig:MSE_vs_itr}
\end{figure}

We compare the empirical MSE performance of the Douglas-Rachford algorithm for~\eqref{eq:L1_optimization} and its prediction obtained from Conjecture~\ref{conj:DR}. 
In Fig.~\ref{fig:MSE_vs_itr}, we show the MSE performance versus the number of iterations in the algorithm, where $\Delta = 0.7$, $p_{0} = 0.9$, and $\sigma_{\txv}^{2} = 0.001$. 
The parameters of the Douglas-Rachford algorithm are set as $\gamma = 10$ and $\rho_{k} = 1$. 
The initial values of $\bm{s}^{(k)}$ and $\bm{z}^{(k)}$ are given by $\bm{s}^{(0)} = \bm{z}^{(0)} = \bm{0}$. 
In Fig.~\ref{fig:MSE_vs_itr}, `empirical' refers to the empirical MSE performance of the Douglas-Rachford algorithm  in~\eqref{eq:update_s_DR} and~\eqref{eq:update_z_DR}. 
The MSE curve is the average of the performance for $500$ independent realizations of the reconstruction problem. 
Also, `prediction' represents the predicted MSE obtained by Conjecture~\ref{conj:DR}. 
To calculate the prediction, we compute $(\alpha_{k}^{\ast}, \beta_{k}^{\ast})$ and $100,000$ realizations of $(S_{k}, Z_{k})$ for $k = 0, 1, \dotsc$ from~\eqref{eq:update_S} and~\eqref{eq:update_Z}, following the similar way to~\cite{hayakawa2022}. 
Here, the expectation in~\eqref{eq:SO_k} is approximated by using the realizations of $(S_{k}, Z_{k})$. 
In Fig.~\ref{fig:MSE_vs_itr}, we also show the asymptotic MSE of the optimizer of~\eqref{eq:L1_optimization} as `asymptotic MSE of optimizer,' which is derived by using the standard CGMT approach~\cite{thrampoulidis2018}. 
The value of the parameter $\lambda$ in~\eqref{eq:L1_optimization} is selected to minimize the asymptotic MSE of the optimizer. 
Figure~\ref{fig:MSE_vs_itr} shows that the empirical performance is close to the prediction when $N$ is sufficiently large. 
We can also see that they approach the asymptotic MSE of the optimizer after sufficient iterations. 
To be precise, however, we can observe a slight difference between the prediction and the empirical performance. 
This is partly because we evaluate the empirical performance for finite $N$, while we assume $M, N \to \infty$ to obtain the asymptotic prediction. 
Another possible reason is that we approximate $(\alpha_{k}^{\ast}, \beta_{k}^{\ast})$ by using many realizations of $(S_{k}, Z_{k})$ in the prediction. 

\begin{figure}[!t]
    \centering
    \includegraphics[width=85mm]{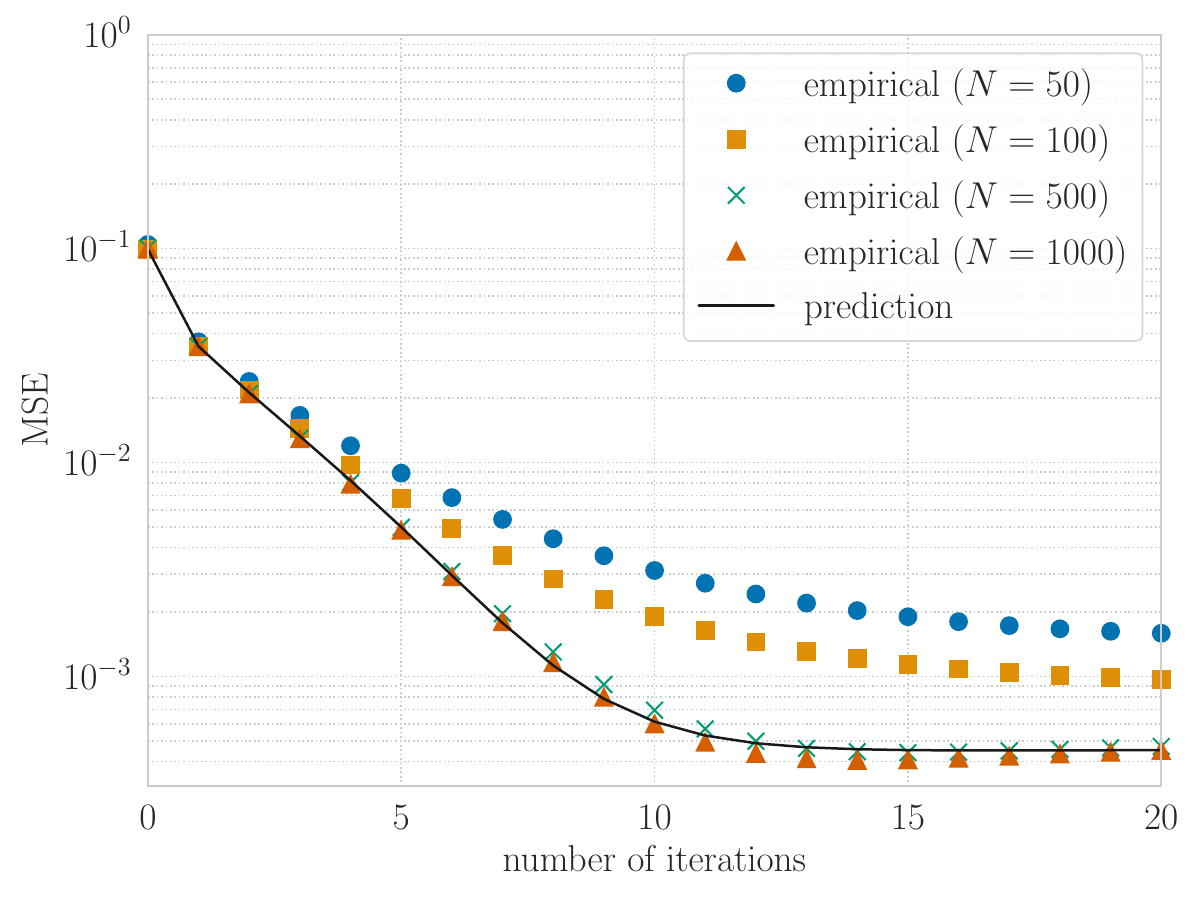}
    \caption{MSE performance of Douglas-Rachford algorithm with elastic net regularization ($\Delta = 0.7$, $p_{0} = 0.9$, $\sigma_{\txv}^{2} = 0.0001$, $\lambda_{1} = \lambda_{2} = 0.01$, $\gamma = 1$, $\rho_{k} = 1$).}
    \label{fig:MSE_vs_itr_EN}
\end{figure}

We also investigate the performance of the Douglas-Rachford algorithm with the elastic net regularization in Fig.~\ref{fig:MSE_vs_itr_EN}. 
In the simulation, we set $\Delta = 0.7$, $p_{0} = 0.9$, $\sigma_{\txv}^{2} = 0.0001$, $\lambda_{1} = \lambda_{2} = 0.01$, $\gamma = 10$, and $\rho_{k} = 1$. 
As in the case with the $\ell_{1}$ regularization, the empirical performance is well predicted by the prediction when $N$ is sufficiently large. 

\begin{figure}[!t]
    \centering
    \includegraphics[width=85mm]{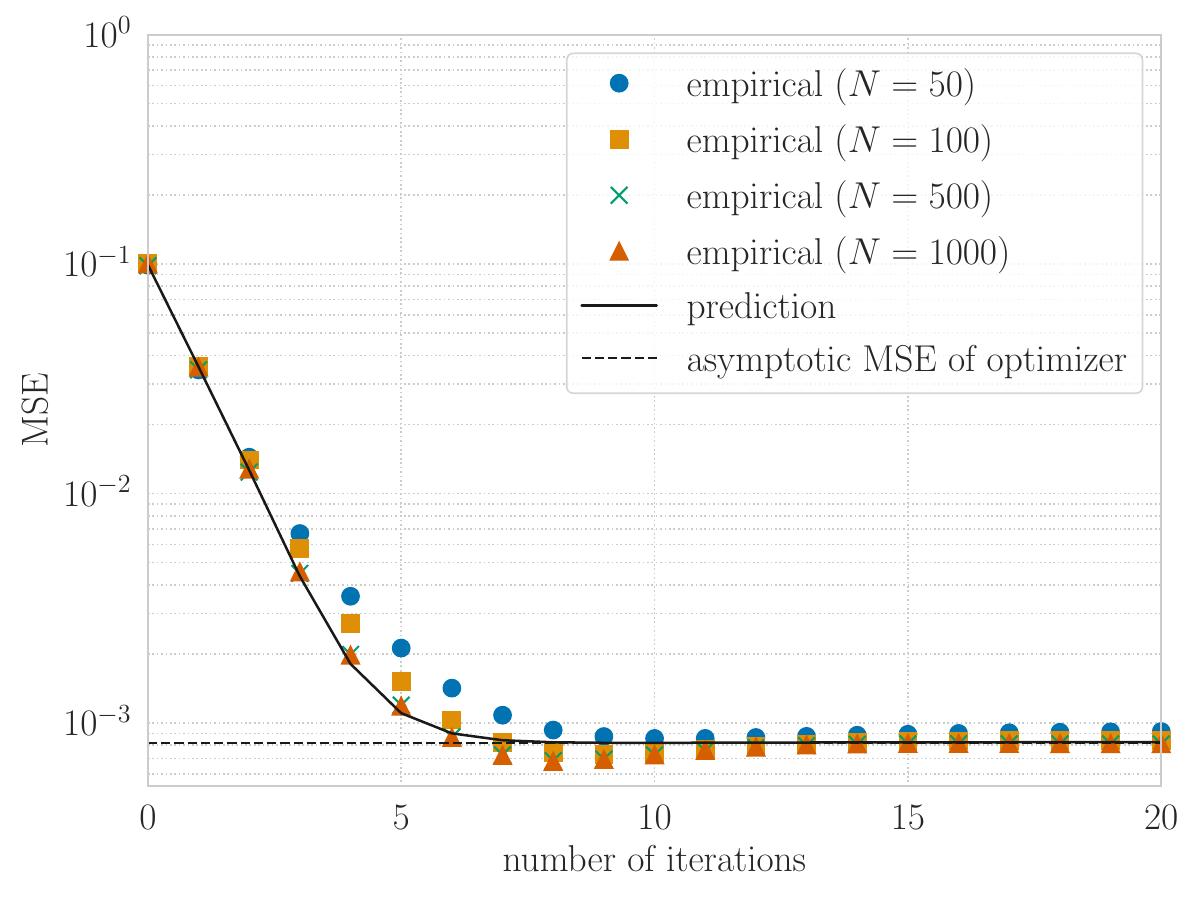}
    \caption{MSE performance of Douglas-Rachford algorithm for $\ell_{1}$ optimization (Bernoulli measurement matrix, $\Delta = 0.7$, $p_{0} = 0.9$, $\sigma_{\txv}^{2} = 0.001$, $\gamma = 10$, $\rho_{k} = 1$).}
    \label{fig:MSE_vs_itr_Bernoulli}
\end{figure}

Fig.~\ref{fig:MSE_vs_itr_Bernoulli} shows the MSE performance of the Douglas-Rachford algorithm with the $\ell_{1}$ regularization for the Bernoulli measurement matrix. 
In the evaluation of the empirical performance, the element of the measurement matrix $\bm{A} \in \mathbb{R}^{M \times N}$ is generated from the i.i.d.\ distribution given by $\Pr (a_{m, n} = 1/\sqrt{N}) = \Pr (a_{m, n} = -1/\sqrt{N}) = 0.5$, where $a_{m, n}$ denotes the $(m, n)$ element of $\bm{A}$. 
For other parameters, we set the same values as in Fig.~\ref{fig:MSE_vs_itr}. 
From Fig.~\ref{fig:MSE_vs_itr_Bernoulli}, we can see that the empirical performance is well predicted by the prediction even for the Bernoulli measurement matrix, though the Gaussian measurement matrix is assumed in derivation of the prediction. 

\begin{figure}[!t]
    \centering
    \includegraphics[width=85mm]{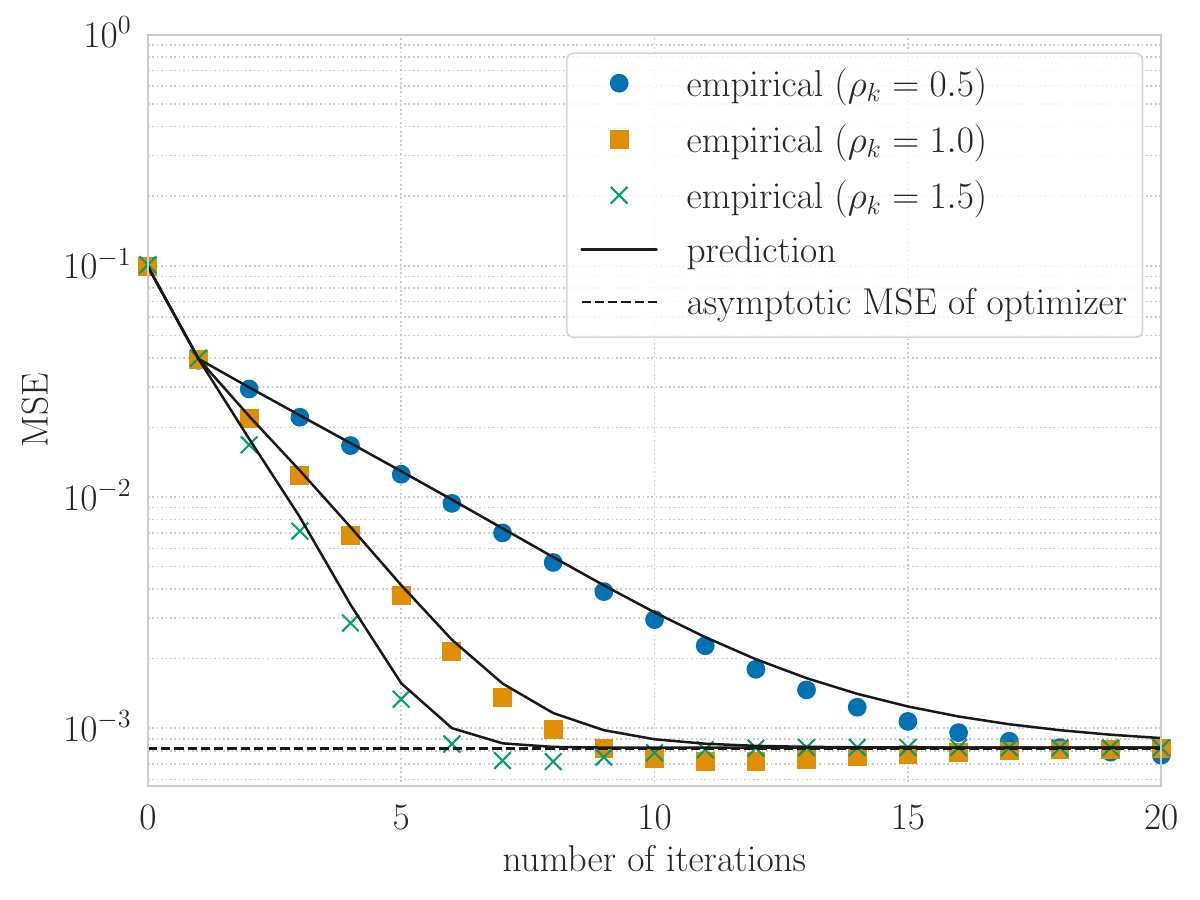}
    \caption{MSE performance of Douglas-Rachford algorithm for $\ell_{1}$ optimization ($N = 1000$, $\Delta = 0.7$, $p_{0} = 0.9$, $\sigma_{\txv}^{2} = 0.001$, $\gamma = 5.0$).}
    \label{fig:MSE_vs_itr_rho}
\end{figure}

Figure~\ref{fig:MSE_vs_itr_rho} shows the MSE performance of the Douglas-Rachford algorithm with the $\ell_{1}$ regularization for different values of $\rho_{k}$.
In the figure, the parameters are $N = 1000$, $\Delta = 0.7$, $p_{0} = 0.9$, $\sigma_{\txv}^{2} = 0.001$, and $\gamma = 5.0$. 
From the figure, we observe that the empirical performance is well predicted for every different values of $\rho_{k}$. 
We can also see that the convergence speed is significantly influenced by the value of $\rho_{k}$. 

\begin{figure}[!t]
    \centering
    \includegraphics[width=85mm]{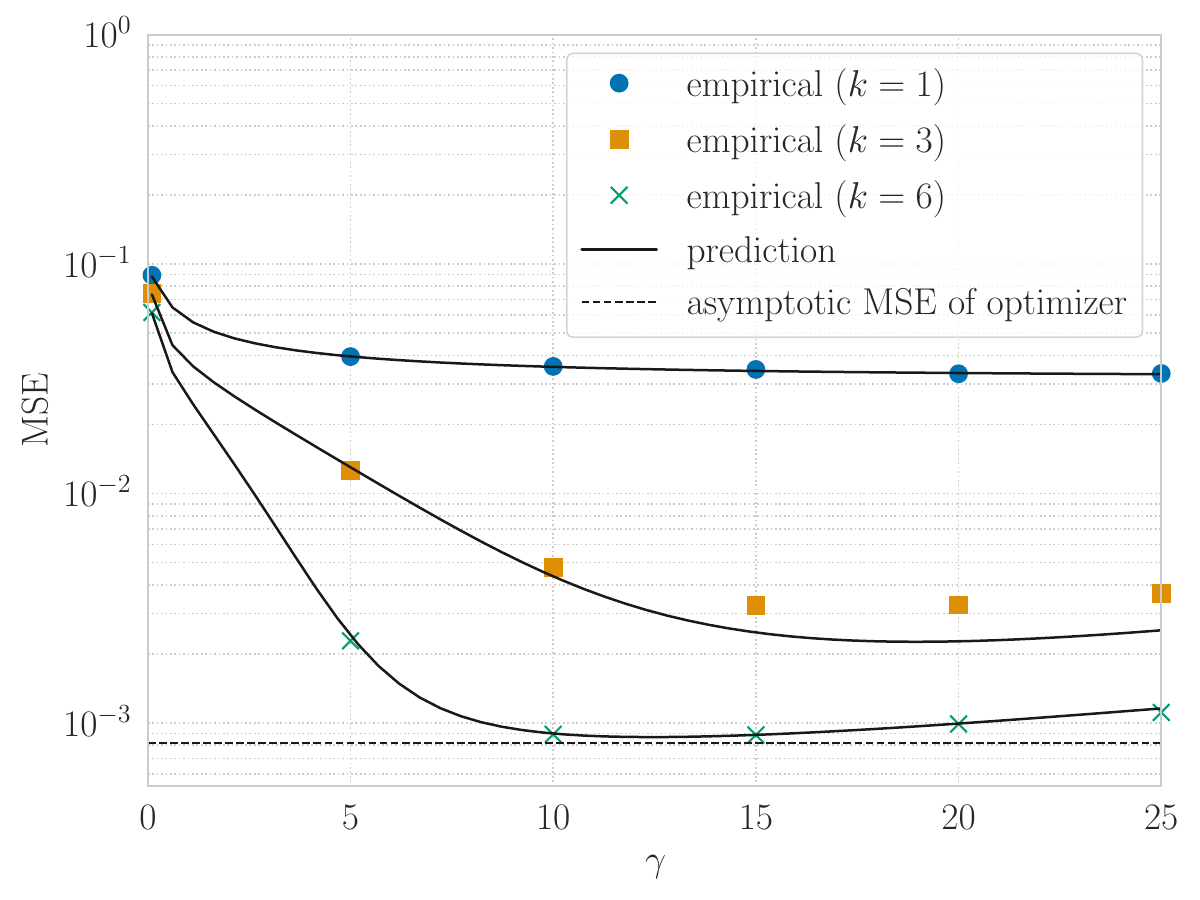}
    \caption{MSE performance of Douglas-Rachford algorithm for $\ell_{1}$ optimization versus $\gamma$ ($N = 500$, $\Delta = 0.7$, $p_{0} = 0.9$, $\sigma_{\txv}^{2} = 0.001$, $\rho_{k} = 1$).}
    \label{fig:MSE_vs_gamma}
\end{figure}

We then show the MSE performance of the Douglas-Rachford algorithm with the $\ell_{1}$ regularization for different parameters $\gamma$ in Fig.~\ref{fig:MSE_vs_gamma}. 
In the figure, we have $N = 500$, $\Delta = 0.7$, $p_{0} = 0.9$, $\sigma_{\txv}^{2} = 0.001$, and $\rho_{k} = 1$. 
From the figure, we can observe that the performance is improved as the iteration index $k$ increases. 
The figure also implies that the value of $\gamma$ significantly affects the performance of the algorithm. 
Since the empirical performance is well predicted for all $\gamma$, we can tune the parameter by using the prediction. 
The figure shows that we should choose $\gamma$ between $10$ and $15$ in this case. 

\begin{figure}[!t]
    \centering
    \includegraphics[width=85mm]{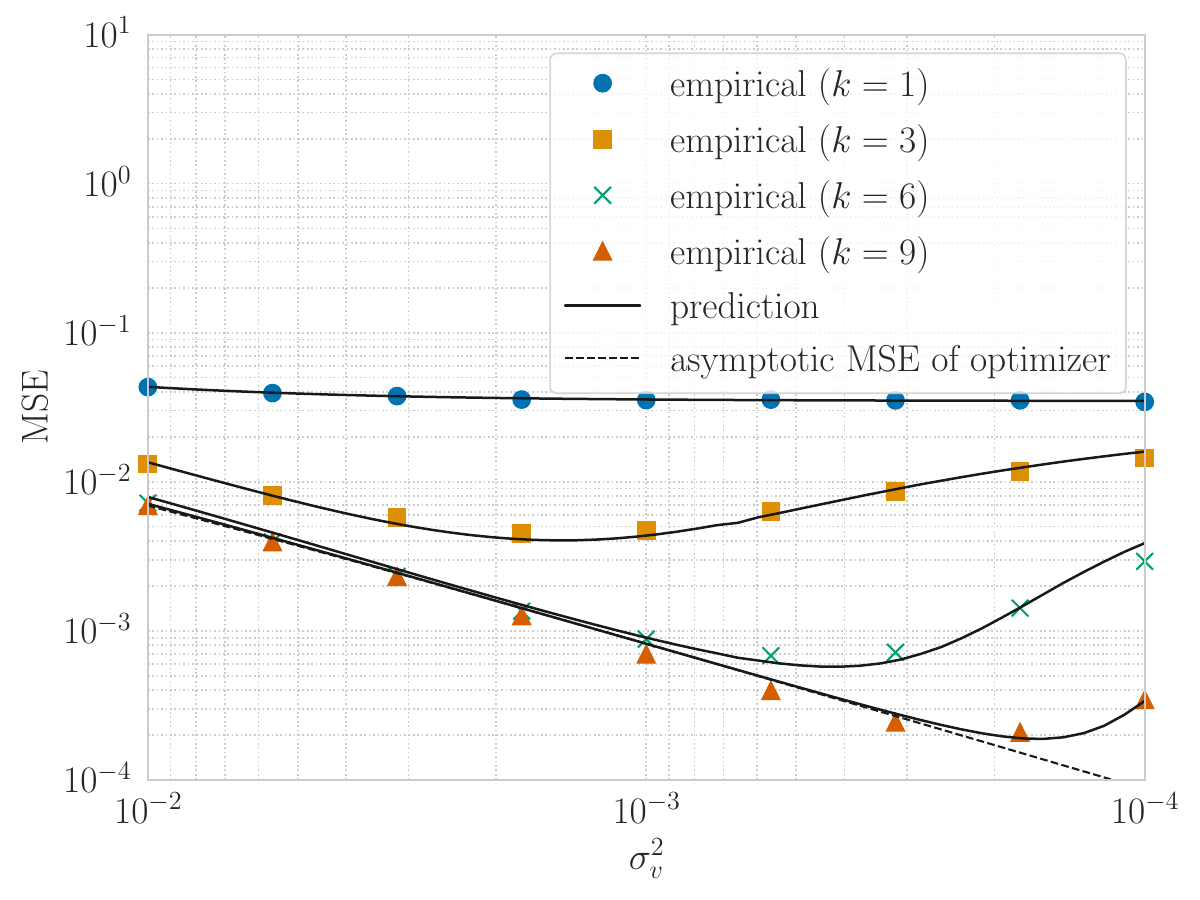}
    \caption{MSE performance of Douglas-Rachford algorithm for $\ell_{1}$ optimization versus $\hat{\sigma}_{\txv}^{2}$ ($N = 500, \Delta = 0.7, p_{0} = 0.9, \gamma = 10, \rho_{k} = 1$).}
    \label{fig:MSE_vs_sigma2}
\end{figure}

We also investigate the performance of the Douglas-Rachford algorithm with the $\ell_{1}$ regularization for different values of $\sigma_{\txv}^{2}$ in Fig.~\ref{fig:MSE_vs_sigma2}. 
In the figure, we set $N = 500$, $\Delta = 0.7$, $p_{0} = 0.9$, $\gamma = 10$, and $\rho_{k} = 1$. 
From Fig.~\ref{fig:MSE_vs_sigma2}, we can see that the empirical performance is well predicted for different values of $\sigma_{\txv}^{2}$. 
The figure also implies that the performance is improved as the value of $\sigma_{\txv}^{2}$ decreases, though the required number of iterations for the convergence increases. 
\subsection{Performance Prediction of Douglas-Rachford Algorithm with Nonconvex Regularization}
\begin{figure}[!t]
    \centering
    \includegraphics[width=85mm]{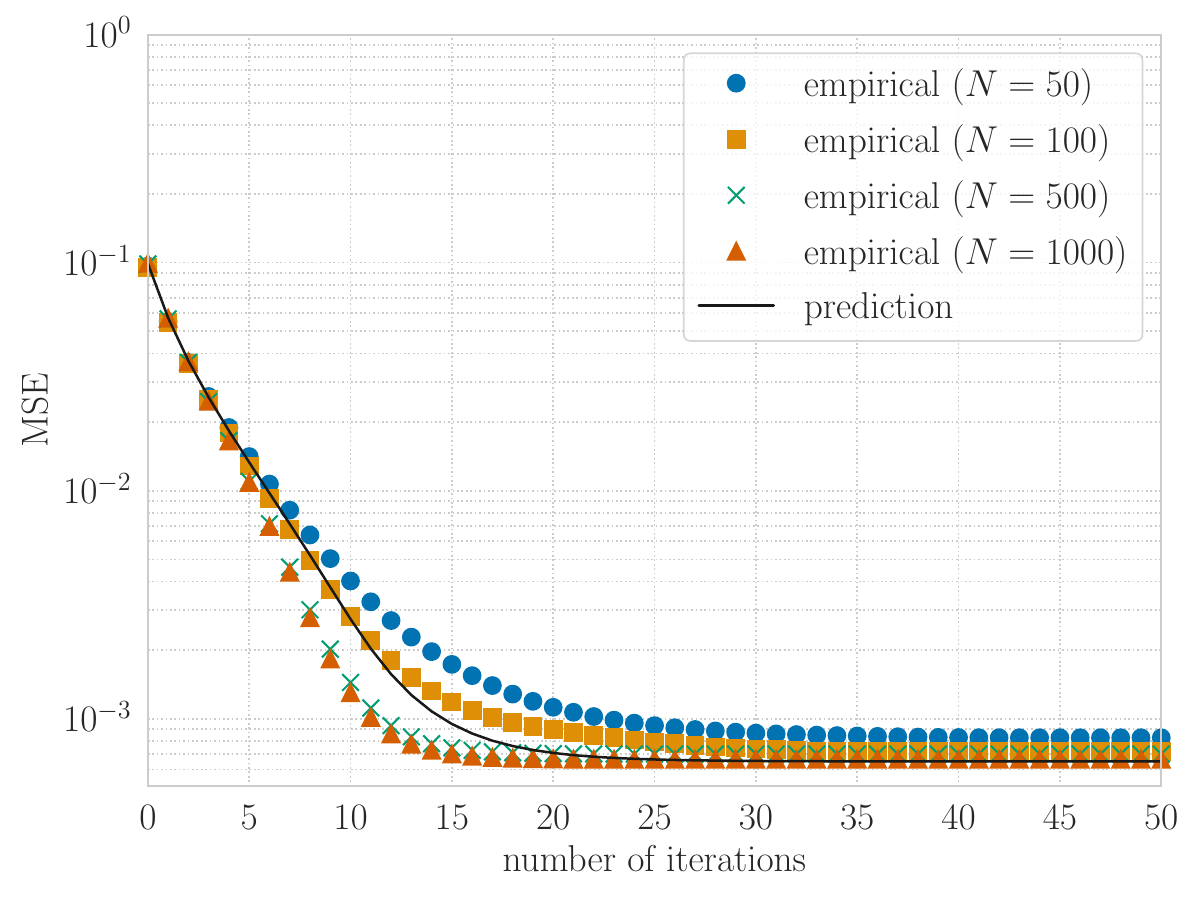}
    \caption{MSE performance of Douglas-Rachford algorithm with nonconvex SCAD regularization ($\Delta = 0.7$, $p_{0} = 0.9$, $\sigma_{\txv}^{2} = 0.001$, $\lambda = 0.1$, $\gamma = 1$, $\rho_{k} = 1$, $a = 4$).}
    \label{fig:MSE_vs_itr_SCAD}
\end{figure}

Finally, we evaluate the performance for nonconvex sparse regularization. 
Figure~\ref{fig:MSE_vs_itr_SCAD} shows the MSE performance of the Douglas-Rachford algorithm with SCAD regularization in~\eqref{eq:SCAD}, where $\Delta = 0.7$, $p_{0} = 0.9$, $\sigma_{\txv}^{2} = 0.001$, $\lambda = 0.1$, $\gamma = 1$, $\rho_{k} = 1$, and $a = 4$. 
The empirical performance and the prediction are obtained in the same way as Fig.~\ref{fig:MSE_vs_itr}. 
From Fig.~\ref{fig:MSE_vs_itr_SCAD}, we observe that the empirical MSE and its prediction converge to almost the same value in this case. 
In the middle of iterations, however, the behavior of the prediction is different from Fig.~\ref{fig:MSE_vs_itr}, and the empirical performance when $N = 500, 1000$ is better than the prediction. 
One of the possible reasons is the nonconvexity of the SCAD regularizer, and further investigation is necessary for a clear understanding. 
To obtain more precise results, we might need to taking the nonconvexity into account in the analysis. 

\begin{figure}[!t]
    \centering
    \includegraphics[width=85mm]{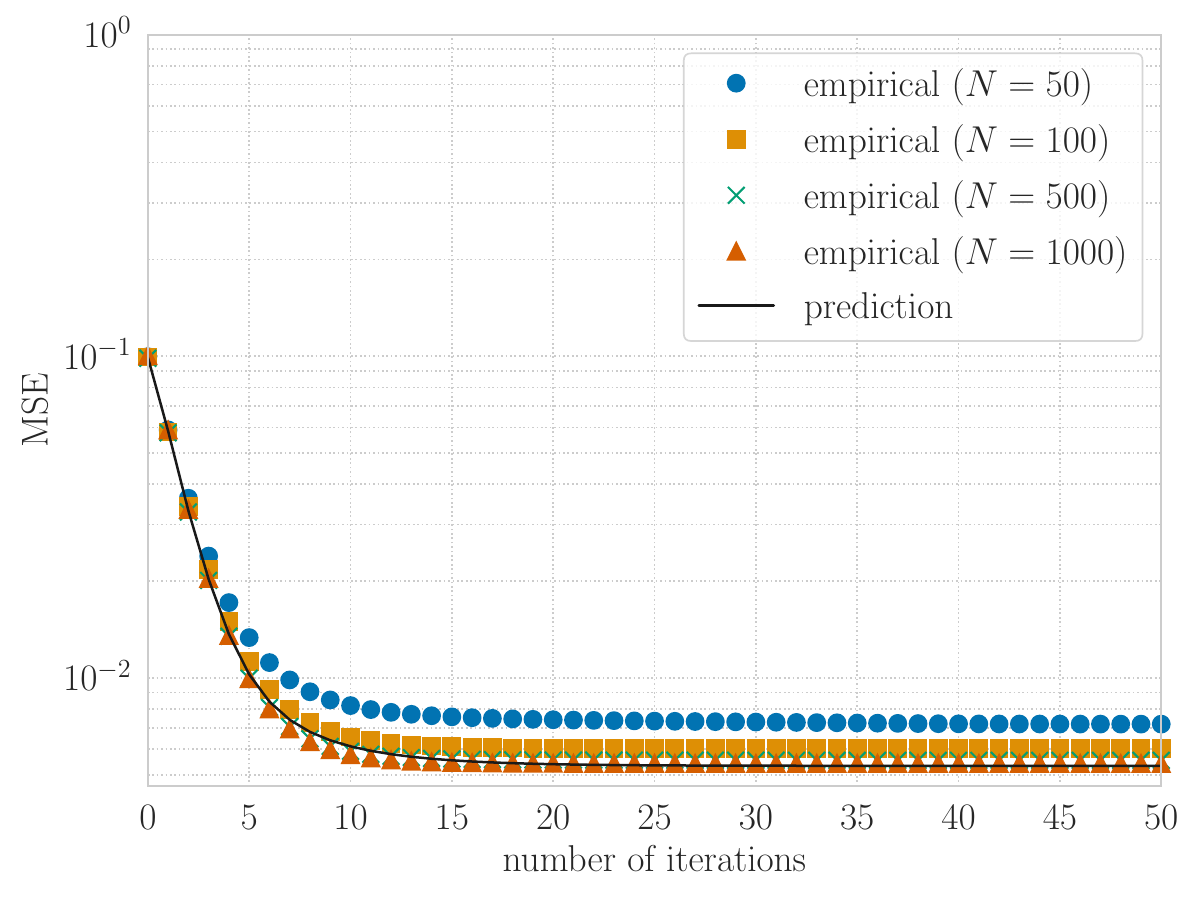}
    \caption{MSE performance of Douglas-Rachford algorithm with SCAD regularization ($\Delta = 0.7, p_{0} = 0.9, \sigma_{\txv}^{2} = 0.01, \lambda = 0.2, \gamma = 1, \rho_{k} = 1, a = 4$).}
    \label{fig:MSE_vs_itr_SCAD_sigma2_0.01}
\end{figure}

We also investigate the performance of the Douglas-Rachford algorithm with SCAD regularization for a larger noise variance $\sigma_{\txv}^{2} = 0.01$ in Fig.~\ref{fig:MSE_vs_itr_SCAD_sigma2_0.01}, where $\Delta = 0.7$, $p_{0} = 0.9$, $\lambda = 0.2$, $\gamma = 1$, $\rho_{k} = 1$, and $a = 4$. 
From the figure, we can see that the empirical performance is well predicted in this case. 
The difference in behavior compared to Fig.~\ref{fig:MSE_vs_itr_SCAD} is probably influenced by the value of $\sigma_{\txv}^{2}$ and other parameters, presenting an intriguing avenue for future research.

\begin{figure}[!t]
    \centering
    \includegraphics[width=85mm]{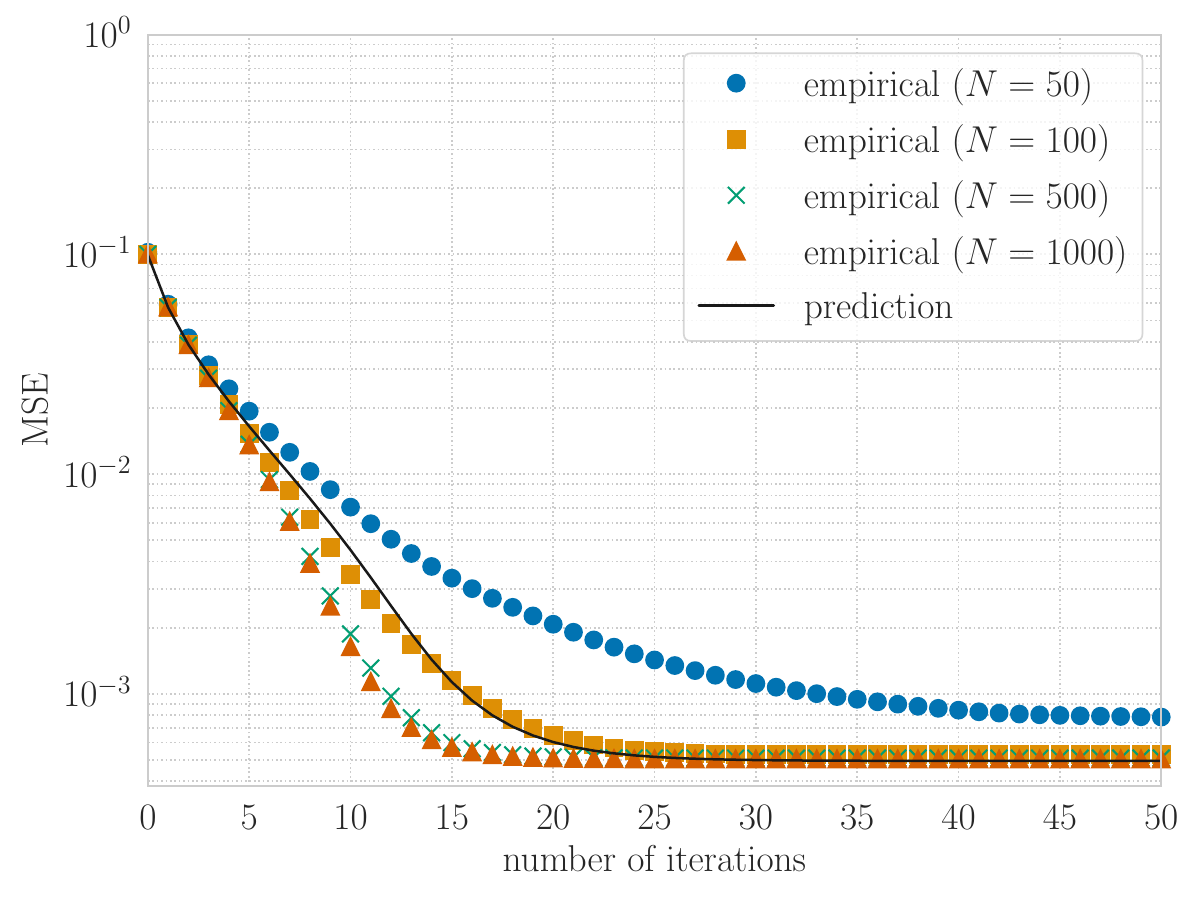}
    \caption{MSE performance of Douglas-Rachford algorithm with nonconvex MCP regularization ($\Delta = 0.7$, $p_{0} = 0.9$, $\sigma_{\txv}^{2} = 0.001$, $\lambda = 0.1$, $\gamma = 1$, $\rho_{k} = 1$, $b = 4$).}
    \label{fig:MSE_vs_itr_MCP}
\end{figure}

Figure~\ref{fig:MSE_vs_itr_MCP} shows the performance of the Douglas-Rachford algorithm with the nonconvex MCP regularization. 
In the simulation, we set $\Delta = 0.7$, $p_{0} = 0.9$, $\sigma_{\txv}^{2} = 0.001$, $\lambda = 0.1$, $\gamma = 1$, $\rho_{k} = 1$, and $b = 4$. 
As in Fig.~\ref{fig:MSE_vs_itr_SCAD}, the empirical performance when $N = 500, 1000$ is better than the prediction in the middle of iterations. 
These results suggest that the nonconvexity of the MCP regularizer also affects the behavior of the prediction. 

\begin{figure}[!t]
    \centering
    \includegraphics[width=85mm]{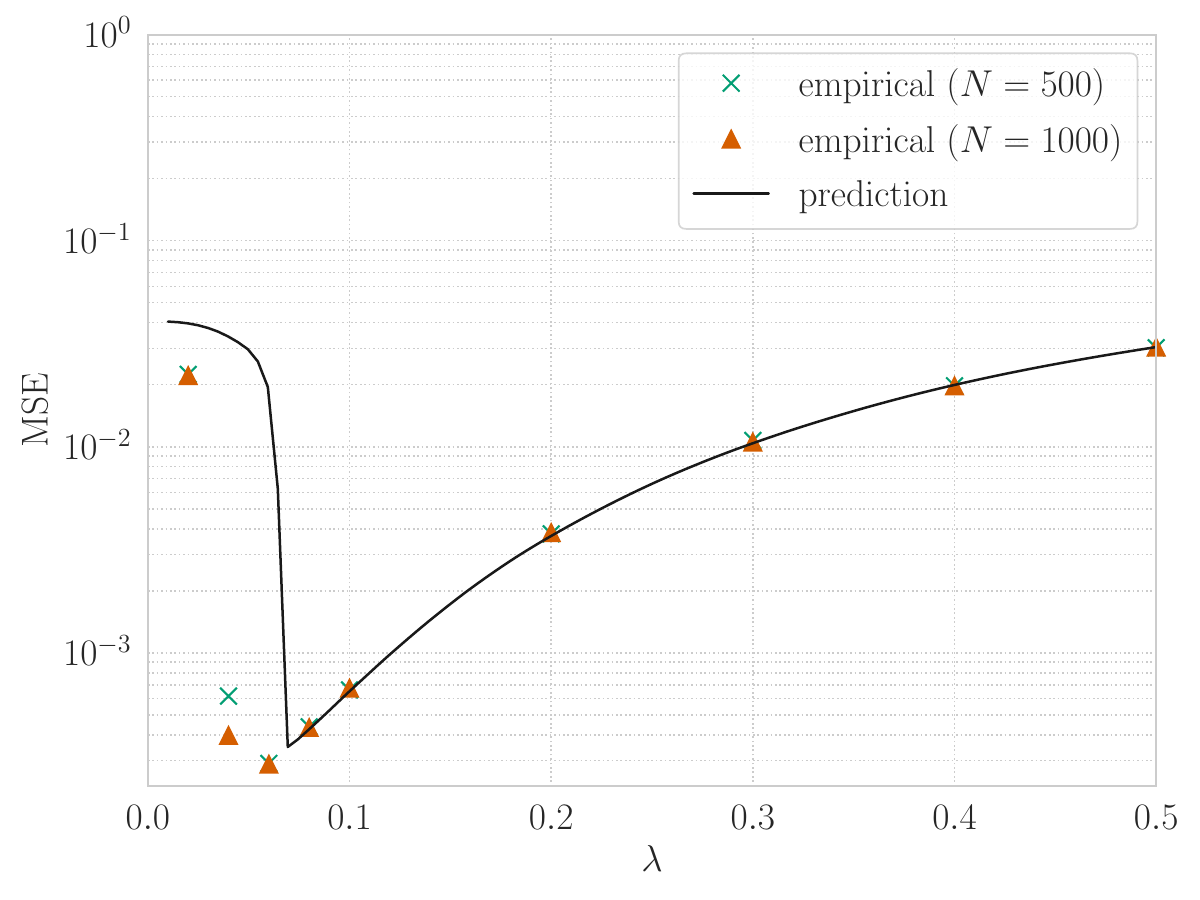}
    \caption{MSE performance of Douglas-Rachford algorithm with nonconvex SCAD regularization versus $\lambda$ ($\Delta = 0.7$, $p_{0} = 0.9$, $\sigma_{\txv}^{2} = 0.001$,  $\gamma = 1$, $\rho_{k} = 1$, $a = 4$, $k = 50$).}
    \label{fig:MSE_vs_lmd_SCAD}
\end{figure}

Figure~\ref{fig:MSE_vs_lmd_SCAD} shows the MSE at the $50$-th iteration versus the regularization parameter $\lambda$, where $\Delta = 0.7$, $p_{0} = 0.9$, $\sigma_{\txv}^{2} = 0.001$,  $\gamma = 1$, and $\rho_{k} = 1$. 
In the simulation, we use the SCAD regularization with $a = 4$. 
From the figure, we can see that the prediction is close to the empirical performance for $\lambda \ge 0.08$, whereas the prediction is worse for smaller values of $\lambda$. 
Even in this case, however, we can tune a reasonable value of $\lambda \approx 0.07$ on the basis of the prediction, without the empirical reconstruction. 
%
\section{Conclusion} \label{sec:conclusion}
In this paper, we have analyzed the asymptotic behavior of the proximity operator for the squared loss function in linear inverse problems. 
We have also shown that the result in Proposition~\ref{prop:prox} can be used for the asymptotic performance prediction of some optimization algorithms for large-scale linear inverse problems. 
Simulation results show that the empirical distribution of the output of the proximity operator agrees well with the theoretical prediction by Proposition~\ref{prop:prox}. 
Moreover, the empirical MSE performance of the Douglas-Rachford algorithm can be well predicted in large-scale compressed sensing via the $\ell_{1}$ optimization. 
For the nonconvex SCAD regularization and MCP regularization, the prediction is also valid to some extent, although the behavior is slightly different from the convex case. 
These results suggest that the proposed prediction method can be a foundation of the rigorous analysis of optimization algorithms for large-scale linear inverse problems.

Future work includes the rigorous analysis of the prediction by Conjecture~\ref{conj:DR} and the further investigation of interesting behavior for the case with nonconvex regularization. 
One of possible directions is to extend or modify the approach in~\cite{chandrasekher2023}, though independent measurement matrices at each iteration of the algorithm is assumed in their analysis. 
Extensions of our approach beyond Assumptions~\ref{ass:distribution} and~\ref{ass:regularizer} would also be an important research direction. 
Applications of our results to optimization algorithms other than the Douglas-Rachford algorithm would also be an interesting topic. 
\appendices
\section{Derivation of~\eqref{eq:SO}} \label{app:proof}
We provide an overview of the derivation of the optimization problem in~\eqref{eq:SO} from~\eqref{eq:prox_L_def}. 
Since the procedure is almost the same as the CGMT-based analyses~\cite{thrampoulidis2018, hayakawa2020a, hayakawa2022}, the rigorous discussion is omitted in some parts. 

We first define $\bm{e} = \bm{s}-\bm{x}$. 
The problem in~\eqref{eq:prox_L_def} can be rewritten as the optimization problem for $\bm{e}$, i.e., 
\begin{align}
    \min_{\bm{e} \in \mathbb{R}^{N}} 
    \frac{1}{N} 
    \curbra{
        \frac{1}{2} \norm{\bm{A}\bm{e}-\bm{v}}_{2}^{2} + \frac{1}{2 \gamma} \norm{ \bm{e} + \bm{x} - \bm{z}}_{2}^{2} 
    }. 
\end{align}
Note that we have normalized the objective function by $N$. 
From the property 
\begin{align}
    \frac{1}{2} \norm{\bm{A} \bm{e} - \bm{v}}_{2}^{2} = 
    \max_{\bm{u} \in \mathbb{R}^{M}} 
    \curbra{
        \sqrt{N} \bm{u}^{\top} \paren{\bm{A} \bm{e} - \bm{v}} - \frac{N}{2} \norm{\bm{u}}_{2}^{2} 
    }, 
\end{align}
we can obtain the optimization problem 
\begin{align}
    &\min_{\bm{e} \in \mathbb{R}^{N}} \max_{\bm{u} \in \mathbb{R}^{M}} 
    \Biggl\{
        \frac{1}{N} \bm{u}^{\top} \paren{\sqrt{N} \bm{A}} \bm{e} 
        - \frac{1}{\sqrt{N}} \bm{v}^{\top}\bm{u} \notag \\\
    &\hspace{25mm}
        - \frac{1}{2} \norm{\bm{u}}_{2}^{2}
        + \frac{1}{N} \frac{1}{2 \gamma} \norm{ \bm{e} + \bm{x} - \bm{z} }_{2}^{2} 
    \Biggr\}. \label{eq:PO}
\end{align}

CGMT~\cite{thrampoulidis2018} enables us to analyze the optimization problem 
\begin{align}
    &\min_{\bm{e} \in \mathbb{R}^{N}} \max_{\bm{u} \in \mathbb{R}^{M}}
    \Biggl\{
        \frac{1}{N} \paren{\norm{\bm{e}}_{2} \bm{g}^{\top} \bm{u} - \norm{\bm{u}}_{2} \bm{h}^{\top} \bm{e}} 
        - \frac{1}{\sqrt{N}} \bm{v}^{\top}\bm{u}  \notag \\
    &\hspace{25mm}
        - \frac{1}{2} \norm{\bm{u}}_{2}^{2}
        + \frac{1}{N} \frac{1}{2 \gamma} \norm{ \bm{e} + \bm{x} - \bm{z} }_{2}^{2} 
    \Biggr\} \label{eq:AO}
\end{align}
instead of~\eqref{eq:PO}, where the elements of $\bm{g} \in \mathbb{R}^{M}$ and $\bm{h} \in \mathbb{R}^{N}$ are i.i.d.\ random variables with $\mathcal{N}(0, 1)$. 
Considering that both $\bm{g}$ and $\bm{v}$ are Gaussian, each element $\frac{\norm{\bm{e}}_{2}}{\sqrt{N}}\bm{g} - \bm{v}$ is also Gaussian with $\mathcal{N} \paren{0, \frac{\norm{\bm{e}}_{2}^{2}}{N} + \sigma_{\txv}^{2}}$. 
We can thus simplify $\paren{\frac{\norm{\bm{e}}_{2}}{\sqrt{N}}\bm{g}-\bm{v}}^{\top} \bm{u}$ as $\sqrt{\frac{\norm{\bm{e}}_{2}^{2}}{N} + \sigma_{\txv}^{2}} \bm{g}^{\top} \bm{u}$, where we use the notation $\bm{g}$ to represent a vector with the elements following $\mathcal{N}(0, 1)$. 
Hence, we can rewrite~\eqref{eq:AO} as 
\begin{align}
    &\min_{\bm{e} \in \mathbb{R}^{N}} \max_{\bm{u} \in \mathbb{R}^{M}}
    \Biggl\{
        \frac{1}{\sqrt{N}} \sqrt{\frac{\norm{\bm{e}}_{2}^{2}}{N} + \sigma_{\txv}^{2}} \bm{g}^{\top} \bm{u} 
        - \frac{1}{N} \norm{\bm{u}}_{2} \bm{h}^{\top} \bm{e} \notag \\
    &\hspace{25mm}
        - \frac{1}{2} \norm{\bm{u}}_{2}^{2}
        + \frac{1}{N} \frac{1}{2 \gamma} \norm{ \bm{e} + \bm{x} - \bm{z} }_{2}^{2} 
    \Biggr\}. \label{eq:AO_supplement1}
\end{align}
If we define $\beta = \norm{\bm{u}}_{2}$, the maximum value of $\bm{g}^{\top} \bm{u}$ in the first term of~\eqref{eq:AO_supplement1} can be written as $\beta \norm{\bm{g}}_{2}$. 
Moreover, we use 
\begin{align}
    \sqrt{\frac{\norm{\bm{e}}_{2}^{2}}{N} + \sigma_{\txv}^{2}}
    &= 
    \min_{\alpha>0}
    \paren{
        \frac{\alpha}{2} + \frac{\frac{\norm{\bm{e}}_{2}^{2}}{N} + \sigma_{\txv}^{2}}{2\alpha} 
    } \label{eq:root_eliminate_trick}
\end{align}
and rewrite the square root term in~\eqref{eq:AO_supplement1} to obtain 
\begin{align}
    &\min_{\bm{e} \in \mathbb{R}^{N}} \max_{\beta \ge 0} \min_{\alpha>0} 
    \Biggl\{
        \frac{\alpha\beta}{2} \frac{\norm{\bm{g}}_{2}}{\sqrt{N}} 
        + \frac{1}{N} \frac{\beta}{2 \alpha} \frac{\norm{\bm{g}}_{2}}{\sqrt{N}} \norm{\bm{e}}_{2}^{2} + \frac{\beta\sigma_{\txv}^{2}}{2\alpha} \frac{\norm{\bm{g}}_{2}}{\sqrt{N}} \notag \\
    &\hspace{15mm}
        - \frac{\beta}{N} \bm{h}^{\top} \bm{e}
        - \frac{1}{2} \beta^{2}
        + \frac{1}{N} \frac{1}{2 \gamma} \norm{ \bm{e} + \bm{x} - \bm{z} }_{2}^{2} 
    \Biggr\}. \label{eq:AO_supplement3}
\end{align}
We can further rewrite~\eqref{eq:AO_supplement3} as 
\begin{align}
    &\min_{\alpha>0} \max_{\beta \ge 0} 
    \Biggl\{
        \frac{\alpha\beta}{2} \frac{\norm{\bm{g}}_{2}}{\sqrt{N}} 
        + \frac{\beta\sigma_{\txv}^{2}}{2\alpha} \frac{\norm{\bm{g}}_{2}}{\sqrt{N}}
        - \frac{1}{2} \beta^{2} \notag \\
    &\hspace{30mm}
        + \min_{\bm{e} \in \mathbb{R}^{N}} 
        \frac{1}{N} \sum_{n=1}^{N} J_{n}(e_{n}, \alpha, \beta) \label{eq:AO2}
    \Biggr\}, 
\end{align}
where $(\cdot)_{n}$ represents the $n$-th element of the corresponding bold vector and 
\begin{align}
    J_{n} (e_{n}, \alpha, \beta) 
    = 
    \frac{\beta}{2\alpha} \frac{\norm{\bm{g}}_{2}}{\sqrt{N}} e_{n}^{2} 
    - \beta h_{n} e_{n} 
    + \frac{1}{2 \gamma} \paren{ e_{n} + x_{n} - z_{n} }^{2}. 
\end{align}
The minimum value of $J_{n}(e_{n}, \alpha, \beta)$ over $e_{n}$ is given by $J_{n}(\hat{s}_{n}(\alpha, \beta) - x_{n}, \alpha, \beta)$, where 
\begin{align}
    &\hat{s}_{n} (\alpha, \beta) \notag \\
    &=
    \dfrac{1}{\dfrac{\beta}{\alpha}\dfrac{\norm{\bm{g}}_{2}}{\sqrt{N}} + \dfrac{1}{\gamma}} 
    \Biggl(
        \frac{\beta}{\alpha}\frac{\norm{\bm{g}}_{2}}{\sqrt{N}} \paren{x_{n} +  \frac{\sqrt{N}}{\norm{\bm{g}}_{2}} \alpha h_{n}} 
    + \frac{1}{\gamma} z_{n}
    \Biggr). 
\end{align}
Hence, we can rewrite the optimization problem~\eqref{eq:AO2} as 
\begin{align}
    &\min_{\alpha>0} \max_{\beta \ge 0} 
    \Biggl\{
        \frac{\alpha\beta}{2} \frac{\norm{\bm{g}}_{2}}{\sqrt{N}} 
        + \frac{\beta\sigma_{\txv}^{2}}{2\alpha} \frac{\norm{\bm{g}}_{2}}{\sqrt{N}}
        - \frac{1}{2} \beta^{2} \notag \\
    &\hspace{20mm}
        + 
        \frac{1}{N} \sum_{n=1}^{N} J_{n} \paren{\hat{s}_{n} (\alpha, \beta)-x_{n}, \alpha, \beta} \label{eq:AO3}
    \Biggr\}. 
\end{align}
We can show that the objective function of the optimization problem in~\eqref{eq:AO3} converges pointwise to~\eqref{eq:SO} as $M, N \to \infty$. 
$J_{n}(\hat{s}_{n}(\alpha, \beta) - x_{n}, \alpha, \beta)$ and $\hat{s}_{n} (\alpha, \beta)$ correspond to~\eqref{eq:J_alpha_beta} and~\eqref{eq:S_alpha_beta}, respectively. 

Intuitively, from the definition of $\alpha$ in~\eqref{eq:root_eliminate_trick}, the optimal value of $\alpha$ corresponds to $\sqrt{\frac{\norm{\bm{e}}_{2}^{2}}{N} + \sigma_{\txv}^{2}}$, which finally results in~\eqref{eq:MSE_prox}. 
For the precise discussion, see~\cite{thrampoulidis2018,hayakawa2020a}. 
\bibliographystyle{IEEEtran}
\bibliography{myBibTeX}

\begin{thebibliography}{10}
\providecommand{\url}[1]{#1}
\csname url@samestyle\endcsname
\providecommand{\newblock}{\relax}
\providecommand{\bibinfo}[2]{#2}
\providecommand{\BIBentrySTDinterwordspacing}{\spaceskip=0pt\relax}
\providecommand{\BIBentryALTinterwordstretchfactor}{4}
\providecommand{\BIBentryALTinterwordspacing}{\spaceskip=\fontdimen2\font plus
\BIBentryALTinterwordstretchfactor\fontdimen3\font minus
  \fontdimen4\font\relax}
\providecommand{\BIBforeignlanguage}[2]{{%
\expandafter\ifx\csname l@#1\endcsname\relax
\typeout{** WARNING: IEEEtran.bst: No hyphenation pattern has been}%
\typeout{** loaded for the language `#1'. Using the pattern for}%
\typeout{** the default language instead.}%
\else
\language=\csname l@#1\endcsname
\fi
#2}}
\providecommand{\BIBdecl}{\relax}
\BIBdecl

\bibitem{candes2005}
E.~J. Cand{\`e}s and T.~Tao, ``Decoding by linear programming,'' \emph{IEEE
  Transactions on Information Theory}, vol.~51, no.~12, pp. 4203--4215, Dec.
  2005.

\bibitem{candes2006}
E.~J. Cand{\`e}s, J.~Romberg, and T.~Tao, ``Robust uncertainty principles:
  {{Exact}} signal reconstruction from highly incomplete frequency
  information,'' \emph{IEEE Transactions on Information Theory}, vol.~52,
  no.~2, pp. 489--509, Feb. 2006.

\bibitem{donoho2006}
D.~L. Donoho, ``Compressed sensing,'' \emph{IEEE Transactions on Information
  Theory}, vol.~52, no.~4, pp. 1289--1306, Apr. 2006.

\bibitem{candes2008b}
E.~J. Cand{\`e}s, ``The restricted isometry property and its implications for
  compressed sensing,'' \emph{Comptes Rendus Mathematique}, vol. 346, no.~9,
  pp. 589--592, May 2008.

\bibitem{sasahara2017}
H.~Sasahara, K.~Hayashi, and M.~Nagahara, ``Multiuser detection based on
  {{MAP}} estimation with sum-of-absolute-values relaxation,'' \emph{IEEE
  Transactions on Signal Processing}, vol.~65, no.~21, pp. 5621--5634, Nov.
  2017.

\bibitem{hayakawa2017a}
R.~Hayakawa and K.~Hayashi, ``Convex optimization-based signal detection for
  massive overloaded {{MIMO}} systems,'' \emph{IEEE Transactions on Wireless
  Communications}, vol.~16, no.~11, pp. 7080--7091, Nov. 2017.

\bibitem{hayakawa2018a}
------, ``Discreteness-aware decoding for overloaded non-orthogonal {{STBCs}}
  via convex optimization,'' \emph{IEEE Communications Letters}, vol.~22,
  no.~10, pp. 2080--2083, Oct. 2018.

\bibitem{daubechies2004}
I.~Daubechies, M.~Defrise, and C.~D. Mol, ``An iterative thresholding algorithm
  for linear inverse problems with a sparsity constraint,''
  \emph{Communications on Pure and Applied Mathematics}, vol.~57, no.~11, pp.
  1413--1457, 2004.

\bibitem{combettes2005}
P.~L. Combettes and V.~R. Wajs, ``Signal recovery by proximal forward-backward
  splitting,'' \emph{Multiscale Modeling \& Simulation}, vol.~4, no.~4, pp.
  1168--1200, Jan. 2005.

\bibitem{figueiredo2007}
M.~A.~T. Figueiredo, R.~D. Nowak, and S.~J. Wright, ``Gradient projection for
  sparse reconstruction: {{Application}} to compressed sensing and other
  inverse problems,'' \emph{IEEE Journal of Selected Topics in Signal
  Processing}, vol.~1, no.~4, pp. 586--597, Dec. 2007.

\bibitem{beck2009}
A.~Beck and M.~Teboulle, ``A fast iterative shrinkage-thresholding algorithm
  for linear inverse problems,'' \emph{SIAM Journal on Imaging Sciences},
  vol.~2, no.~1, pp. 183--202, Jan. 2009.

\bibitem{gabay1976}
D.~Gabay and B.~Mercier, ``A dual algorithm for the solution of nonlinear
  variational problems via finite element approximation,'' \emph{Computers \&
  Mathematics with Applications}, vol.~2, no.~1, pp. 17--40, Jan. 1976.

\bibitem{boyd2011}
S.~Boyd, N.~Parikh, E.~Chu, B.~Peleato, and J.~Eckstein, ``Distributed
  optimization and statistical learning via the alternating direction method of
  multipliers,'' \emph{Found. Trends Mach. Learn.}, vol.~3, no.~1, pp. 1--122,
  Jan. 2011.

\bibitem{lions1979}
P.~Lions and B.~Mercier, ``Splitting algorithms for the sum of two nonlinear
  operators,'' \emph{SIAM Journal on Numerical Analysis}, vol.~16, no.~6, pp.
  964--979, Dec. 1979.

\bibitem{eckstein1992}
J.~Eckstein and D.~P. Bertsekas, ``On the {{Douglas-Rachford}} splitting method
  and the proximal point algorithm for maximal monotone operators,''
  \emph{Mathematical Programming}, vol.~55, no.~1, pp. 293--318, Apr. 1992.

\bibitem{combettes2011}
P.~L. Combettes and J.-C. Pesquet, ``Proximal splitting methods in signal
  processing,'' in \emph{Fixed-{{Point Algorithms}} for {{Inverse Problems}} in
  {{Science}} and {{Engineering}}}, ser. Springer {{Optimization}} and {{Its
  Applications}}.\hskip 1em plus 0.5em minus 0.4em\relax {New York, NY}:
  {Springer New York}, 2011, vol.~49, pp. 185--212.

\bibitem{combettes2019}
P.~L. Combettes and L.~E. Glaudin, ``Fully proximal splitting algorithms in
  image recovery,'' in \emph{Proc. 27th {{European Signal Processing
  Conference}} ({{EUSIPCO}})}, Sep. 2019, pp. 1--5.

\bibitem{combettes2019a}
------, ``Proximal activation of smooth functions in splitting algorithms for
  convex image recovery,'' \emph{SIAM Journal on Imaging Sciences}, vol.~12,
  no.~4, pp. 1905--1935, Jan. 2019.

\bibitem{donoho2011}
D.~L. Donoho, A.~Maleki, and A.~Montanari, ``The noise-sensitivity phase
  transition in compressed sensing,'' \emph{IEEE Transactions on Information
  Theory}, vol.~57, no.~10, pp. 6920--6941, Oct. 2011.

\bibitem{bayati2012}
M.~Bayati and A.~Montanari, ``The {{LASSO}} risk for {{Gaussian}} matrices,''
  \emph{IEEE Transactions on Information Theory}, vol.~58, no.~4, pp.
  1997--2017, Apr. 2012.

\bibitem{donoho2013}
D.~L. Donoho, I.~Johnstone, and A.~Montanari, ``Accurate prediction of phase
  transitions in compressed sensing via a connection to minimax denoising,''
  \emph{IEEE Transactions on Information Theory}, vol.~59, no.~6, pp.
  3396--3433, Jun. 2013.

\bibitem{thrampoulidis2015}
C.~Thrampoulidis, A.~Panahi, and B.~Hassibi, ``Asymptotically exact error
  analysis for the generalized $\ell_{2}^{2}$-{{LASSO}},'' in \emph{Proc.
  {{IEEE International Symposium}} on {{Information Theory}} ({{ISIT}})}, Jun.
  2015, pp. 2021--2025.

\bibitem{thrampoulidis2018}
C.~Thrampoulidis, E.~Abbasi, and B.~Hassibi, ``Precise error analysis of
  regularized ${M}$-estimators in high dimensions,'' \emph{IEEE Transactions on
  Information Theory}, vol.~64, no.~8, pp. 5592--5628, Aug. 2018.

\bibitem{atitallah2017}
I.~B. Atitallah, C.~Thrampoulidis, A.~Kammoun, T.~Y. {Al-Naffouri}, M.~Alouini,
  and B.~Hassibi, ``The {{BOX-LASSO}} with application to {{GSSK}} modulation
  in massive {{MIMO}} systems,'' in \emph{Proc. {{IEEE International
  Symposium}} on {{Information Theory}} ({{ISIT}})}, Jun. 2017, pp. 1082--1086.

\bibitem{thrampoulidis2018a}
C.~Thrampoulidis, W.~Xu, and B.~Hassibi, ``Symbol error rate performance of
  box-relaxation decoders in massive {{MIMO}},'' \emph{IEEE Transactions on
  Signal Processing}, vol.~66, no.~13, pp. 3377--3392, Jul. 2018.

\bibitem{hayakawa2020a}
R.~Hayakawa and K.~Hayashi, ``Asymptotic performance of discrete-valued vector
  reconstruction via box-constrained optimization with sum of $\ell_{1}$
  regularizers,'' \emph{IEEE Transactions on Signal Processing}, vol.~68, pp.
  4320--4335, 2020.

\bibitem{hayakawa2022}
R.~Hayakawa, ``Asymptotic performance prediction for {{ADMM-based}} compressed
  sensing,'' \emph{IEEE Transactions on Signal Processing}, vol.~70, pp.
  5194--5207, 2022.

\bibitem{tan2001}
P.~H. Tan, L.~K. Rasmussen, and T.~J. Lim, ``Constrained maximum-likelihood
  detection in {{CDMA}},'' \emph{IEEE Transactions on Communications}, vol.~49,
  no.~1, pp. 142--153, Jan. 2001.

\bibitem{nagahara2015}
M.~Nagahara, ``Discrete signal reconstruction by sum of absolute values,''
  \emph{IEEE Signal Processing Letters}, vol.~22, no.~10, pp. 1575--1579, Oct.
  2015.

\bibitem{aissa-el-bey2015}
A.~{A{\"i}ssa-El-Bey}, D.~Pastor, S.~M.~A. Sba{\"i}, and Y.~Fadlallah,
  ``Sparsity-based recovery of finite alphabet solutions to underdetermined
  linear systems,'' \emph{IEEE Transactions on Information Theory}, vol.~61,
  no.~4, pp. 2008--2018, Apr. 2015.

\bibitem{hayakawa2018b}
R.~Hayakawa and K.~Hayashi, ``Reconstruction of complex discrete-valued vector
  via convex optimization with sparse regularizers,'' \emph{IEEE Access},
  vol.~6, pp. 66\,499--66\,512, 2018.

\bibitem{Zou2005-pr}
H.~Zou and T.~Hastie, ``\BIBforeignlanguage{en}{Regularization and variable
  selection via the elastic net},'' \emph{\BIBforeignlanguage{en}{J. R. Stat.
  Soc. Series B Stat. Methodol.}}, vol.~67, no.~2, pp. 301--320, Apr. 2005.

\bibitem{wen2018}
F.~Wen, L.~Chu, P.~Liu, and R.~C. Qiu, ``A survey on nonconvex
  regularization-based sparse and low-rank recovery in signal processing,
  statistics, and machine learning,'' \emph{IEEE Access}, vol.~6, pp.
  69\,883--69\,906, 2018.

\bibitem{fan2001}
J.~Fan and R.~Li, ``Variable selection via nonconcave penalized likelihood and
  its oracle properties,'' \emph{Journal of the American Statistical
  Association}, vol.~96, no. 456, pp. 1348--1360, Dec. 2001.

\bibitem{zhang2010a}
C.-H. Zhang, ``Nearly unbiased variable selection under minimax concave
  penalty,'' \emph{The Annals of Statistics}, vol.~38, no.~2, Apr. 2010.

\bibitem{bayati2015}
M.~Bayati, M.~Lelarge, and A.~Montanari, ``Universality in polytope phase
  transitions and message passing algorithms,'' \emph{Annals of Applied
  Probability}, vol.~25, no.~2, pp. 753--822, Apr. 2015.

\bibitem{panahi2017}
A.~Panahi and B.~Hassibi, ``A universal analysis of large-scale regularized
  least squares solutions,'' in \emph{Proc. {{Advances}} in {{Neural
  Information Processing Systems}}}, 2017, pp. 3381--3390.

\bibitem{oymak2018}
S.~Oymak and J.~A. Tropp, ``Universality laws for randomized dimension
  reduction, with applications,'' \emph{Information and Inference: A Journal of
  the IMA}, vol.~7, no.~3, pp. 337--446, Sep. 2018.

\bibitem{luenberger2008}
D.~G. Luenberger and Y.~Ye, ``Basic {{Descent Methods}},'' in \emph{Linear and
  {{Nonlinear Programming}}}, ser. International {{Series}} in {{Operations
  Research}} \& {{Management Science}}.\hskip 1em plus 0.5em minus 0.4em\relax
  {New York, NY}: {Springer US}, 2008, pp. 215--262.

\bibitem{chandrasekher2023}
K.~A. Chandrasekher, A.~Pananjady, and C.~Thrampoulidis, ``Sharp global
  convergence guarantees for iterative nonconvex optimization with random
  data,'' \emph{Ann. Stat.}, vol.~51, no.~1, pp. 179--210, Feb. 2023.

\end{thebibliography}
\begin{IEEEbiographynophoto}{Ryo Hayakawa}
    received the bachelor's degree in engineering, the master's degree in informatics, and Ph.D.\ degree in informatics from Kyoto University, Kyoto, Japan, in 2015, 2017, and 2020, respectively. 
    He is currently an Assistant Professor at Graduate School of Engineering Science, Osaka University. 
    He was a Research Fellow (DC1) of the Japan Society for the Promotion of Science (JSPS) from 2017 to 2020. 
    From 2023, he is an Associate Editor of IEICE Transactions on Fundamentals of Electronics, Communications and Computer Sciences. 
    He received the 33rd Telecom System Technology Student Award, APSIPA ASC 2019 Best Special Session Paper Nomination Award, and the 16th IEEE Kansai Section Student Paper Award. 
    His research interests include signal processing and mathematical optimization. 
    He is a member of IEEE and IEICE. 
\end{IEEEbiographynophoto}
\vfill
\end{document}